\documentclass[aps,prb,twocolumn,longbibliography]{revtex4-2}
\usepackage{amsmath,amssymb,bm,chemformula,epsfig,graphicx,longtable,physics}
\usepackage[utf8]{inputenc}
\usepackage{hyperref}
\usepackage{microtype}
\usepackage{siunitx}
\usepackage{multirow}

\begin{document}

\title{Implementation of self-consistent MGGA functionals in augmented
  plane wave based methods}
\author{Jan Doumont}
\affiliation{Institute of Materials Chemistry, Vienna University of Technology,
Getreidemarkt 9/165-TC, A-1060 Vienna, Austria}
\author{Fabien Tran}
\affiliation{Institute of Materials Chemistry, Vienna University of Technology,
Getreidemarkt 9/165-TC, A-1060 Vienna, Austria}
\author{Peter Blaha}
\affiliation{Institute of Materials Chemistry, Vienna University of Technology,
Getreidemarkt 9/165-TC, A-1060 Vienna, Austria}

\begin{abstract}

  Functionals of the meta-generalized gradient approximation (MGGA)
  are nowadays widely used in chemistry and solid-state physics for
  the simulation of electronic systems like molecules, solids, or
  surfaces. Due to their dependency on the kinetic energy density,
  they are in principle more accurate than GGA functionals for various
  properties (geometry, binding energy, electronic structure, etc.),
  while being nearly as fast since they are still of the semilocal
  form. Thus, when an accuracy better than GGA is required, one may
  consider using a MGGA instead of the much more costly hybrid
  functionals or methods like the random-phase approximation or
  $GW$. In this work, the self-consistent implementation of MGGA
  functionals in APW based methods is presented. Technical aspects of
  the implementation are discussed, and calculations of band gaps,
  lattice constants, and magnetic moments are presented in order to
  validate our implementation. To test the changes of the electron
  density due to a MGGA, the electric field gradient on
  transition-metal atoms is calculated.

\end{abstract}

\maketitle

\section{Introduction}
\label{sec:introduction}

Kohn-Sham (KS) density functional theory (DFT)
\cite{HohenbergPR64,KohnPR65} is, in principle, an exact
theory. However, in practice the exchange-correlation (xc) term in the
total energy functional, $E_{xc}$, is treated approximately since an
usable exact expression for the xc term has not been (and probably
will never be) developed. Various classes of approximations for the xc
energy functional $E_{xc}$ and xc potential $v_{xc}$ exist, and most
of them belong to one of the rungs of Jacob's ladder of DFT
\cite{PerdewAIP01,PerdewJCP05}. Several hundreds of approximations
have been proposed so far
\cite{MarquesCPC12,LehtolaSX18,MardirossianMP17}, and which one to
choose for the problem at hand is not always obvious. The most simple
approximation (first rung of Jacob's ladder) for $E_{xc}$ is the local
density approximation (LDA) \cite{KohnPR65,VoskoCJP80,PerdewPRB92a},
where the xc energy density $\epsilon_{xc}$ is a purely local
functional of the electron density $\rho$,
$E_{xc} = \int \epsilon_{xc} \left( \rho \left( \mathbf{r} \right)
\right) d^3\mathbf{r}$. At the second rung of Jacob's ladder, there is
the generalized gradient approximation (GGA)
\cite{BeckePRA88,PerdewPRL96}, where $\epsilon_{xc}$ depends not only
on the electron density $\rho$, but also on its first derivative
$\nabla\rho$. Functionals of the meta-GGA (MGGA) type
\cite{VanVoorhisJCP98,TaoPRL03,DellaSalaIJQC16}, at the third rung of
Jacob's ladder, depend also on the non-interacting electronic kinetic
energy density (KED) $\tau$ and/or the Laplacian of the electron
density $\nabla^2\rho$. At the fourth rung there are the functionals
using the exact Hartree-Fock (HF) exchange, as the hybrid functionals
\cite{BeckeJCP93b,ErnzerhofJCP99,AdamoJCP99}. Finally, the functionals
at the fifth rung use also the unoccupied Kohn-Sham orbitals (e.g.,
the random-phase approximation \cite{PinesPR52,LangrethSS75}).

Climbing up Jacob's ladder leads to functionals that should, in
principle, be more accurate, but also more complicated to implement
and computationally more expensive. Actually, the functionals of the
fourth and fifth rungs are much more demanding in terms of
computational time and memory. The functionals of the second and third
rungs, the so-called semilocal functionals, are the most widely used,
especially in solid-state physics, where the GGA functionals have been
the standard since the 90s \cite{PerdewPRB92b,PerdewPRL96}. The last
decade has seen a significant increase in popularity of MGGA
\cite{SunPRL15} and hybrid \cite{HeydJCP03} functionals. MGGA
functionals are universally more accurate than GGAs; a MGGA can be
quite accurate for both molecules and solids at the same time, while
this is not possible with any GGA
\cite{PerdewPRL08,ZhaoJCP08,PerdewPRL09,FabianoPRB10,HaasPRB11}. Hybrid
functionals are particularly interesting for properties related to the
electronic structure, like the band gap
\cite{BredowPRB00,MuscatCPL01,PerryPRB01,HeydJCP05}.

The focus of the present work is on MGGA functionals. More
specifically, the self-consistent implementation of KED-dependent MGGA
functionals in augmented plane wave (APW) based methods will be
presented. For this purpose we made use of the WIEN2k code
\cite{WIEN2k,BlahaJCP20}. MGGA functionals have been implemented in a
certain number of codes (see Refs.~
\cite{NeumannMP96,VanVoorhisJCP98,AdamoJCP00,ArbuznikovPCCP02,ArbuznikovCPL03,GrafensteinJCP07,SunPRB11b,FerrighiJCP11,ZaharievJCP13,EichJCP14,YangPRB16,WomackJCP16,YaoJCP17,ReiterJCTC18,YamamotoJCP19}
for works reporting implementation details). However, in the
literature there is no report of the self-consistent implementation of
MGGAs within the APW method \cite{AndersenPRB75,Singh,KarsaiCPC17}. We
note that such an implementation is available in the Elk APW code
\cite{elk}. The approach used in the Elk code differs significantly
from ours, however. In their approach the MGGA potential is added in a
second-variational step, whereas in our approach it is added in the
first variation.

Here, our goal is to derive the novel equations that
arise from the KED-dependency in APW based methods, and to show that
they are correctly implemented. We discuss the results for the band
gap, lattice constant, and magnetic moment. As an application of MGGA
functionals, the electric field gradient (EFG) on transition-metal
atoms is calculated.~\cite{BlahaPRB88}

The paper is organized as follows. Sec.~\ref{sec:theory} gives
details about the theory. Then, validation tests of the implementation
are presented in Sec.~\ref{sec:validation}, while
Sec.~\ref{sec:latticeconstant} discusses the effect of
self-consistency on the lattice constant. As an application of MGGA
functionals, Sec.~\ref{sec:EFG} presents the results for the EFG in
systems with transition-metal atoms. The EFG is especially suited as a
benchmark because of its sensitive dependence on the electronic
density $\rho$. Finally, Sec.~\ref{sec:summary} summarizes the main
points of this work.

\section{Theory}
\label{sec:theory}

\subsection{APW based methods}
\label{sec:APW}

In the all-electron full-potential APW based methods, the unit cell is
partitioned into two disjunct regions: non-overlapping atomic spheres
centered at the nuclei with radii $R_{\mathrm{at}}$ and the
interstitial region. Inside the spheres the orbitals, and all
associated quantities (electron density, potential, \ldots{}), are
expanded in spherical harmonics $Y_{\ell m}$; in the interstitial
region they are expanded in plane waves.

Within the spheres a second partitioning is performed in low-energy
core states and higher energy valence (and semi-core) states. The
former are fully confined to the spheres (they have a vanishing wave
function at and beyond the sphere boundary) and are obtained as
atomic-like solutions of the Dirac equation with a (spherical)
effective KS potential
\cite{desclauxHartreeFockSlater1970a,DesclauxCPC75}. In this case the
variables can be separated in a radial and angular parts in the usual
way \cite{grantRelativisticCalculationAtomic1970}. The core orbitals
are thus given by radial functions multiplied by the spinor spherical
harmonics:
\begin{equation}
\label{eq:psicore}
  \psi^{\alpha}_{n \kappa m} \left( \mathbf{r} \right) =
\begin{pmatrix}
  g_{n \kappa m} \left( r \right) \xi_{\kappa m} \\

  i f_{n \kappa m} \left( r \right) \xi_{-\kappa m}
\end{pmatrix}
\end{equation}
where $n$, $\kappa$, and $m$ are the principal, relativistic, and
magnetic quantum numbers, respectively. $g_{n \kappa m}$ and
$f_{n \kappa m}$ are radial functions (large and small components,
respectively) and $\xi_{\kappa m}$ are the spinor spherical
harmonics. The core states are calculated self-consistently,
i.e. recalculated at each iteration of the self-consistency cycle. For
spin-polarized cases, the effective potential has two components
$v_{\mathrm{KS}} = \begin{pmatrix}v_{\mathrm{KS}}^{\uparrow} &&
  v_{\mathrm{KS}}^{\downarrow}\end{pmatrix}$, such that the
relativistic single-electron wave-function has four components
(large and small components for each spin). Spin mixing of the
  relativistic states is neglected. Note that the contribution
from the core states to the KED is still defined from the
positive-definite form of the non-relativistic kinetic-energy operator
\cite{YePRB15}. This form must be used because (non-relativistic) MGGA
xc functionals are parametrized in terms of the non-relativistic
KED. For example, many MGGA functionals use the iso-orbital indicator
$\alpha_{\text{iso}} = \left(\tau -
  \tau^{\mathrm{W}}\right)/\tau^{\mathrm{TF}}$ (where
$\tau^{\mathrm{W}} = \left\vert\nabla
  \rho\right\vert/\left(8\rho\right)$ and
$\tau^{\mathrm{TF}} = \left(3/10\right)\left( 3\pi^2
\right)^{2/3}\rho^{5/3}$ are the von Weizsäcker
\cite{vonWeizsackerZP35} and Thomas-Fermi
\cite{ThomasPCPS27,FermiRANL27} KED, respectively), assuming it has a
lower bound of zero. For the non-relativistic case this is a safe
assumption, because the KED
$\tau = \tfrac{1}{2} \sum\limits_i \nabla \psi_i^{*} \cdot \nabla \psi_i$ is
positive-definite. For the relativistic KED associated with the
operator $c \boldsymbol{\alpha} \cdot \mathbf{\hat{p}}$
\cite{maierRestoringIsoorbitalLimit2019,grantRelativisticCalculationAtomic1970},
this is not guaranteed and this can lead to wrong results.

The valence states do extend over the whole unit cell and are
described by (linearized) augmented plane waves ((L)APW), depending on
the case. As an illustration, the case of LAPW can trivially be
generalized to APW basis functions, or functions describing semi-core
states (APW+lo, (high-derivative) local orbitals (HDLOs), \ldots{})
\cite{Singh,SjostedtSSC00,MadsenPRB01,MichalicekCPC13,KarsaiCPC17}.

The LAPW basis functions are given by
\begin{align}
  \label{eq:LAPW}
\phi_{\mathbf{K}}\left( \mathbf{r} \right) & =
     \begin{cases}
       \sum\limits_{\ell m} \left[ A^{\mathbf{K}}_{\alpha \ell m} u_{\alpha \ell} \left( r_{\alpha}, E_{\alpha\ell} \right) + \right. \\
          \qquad \left. B^{\mathbf{K}}_{\alpha \ell m} \dot{u}_{\alpha \ell} \left( r_{\alpha}, E_{\alpha\ell} \right) \right] {Y}_{\ell m}    & \mathbf{r} \in \text{S}_{\alpha} \\
           \frac{1}{\sqrt{\Omega}} e^{i \mathbf{K} \cdot \mathbf{r}} & \mathbf{r} \in \text{I},
     \end{cases}
\end{align}
where $\mathbf{K} = \mathbf{k} + \mathbf{G}$ is the sum of the wave
vector $\mathbf{k}$ and the reciprocal lattice vector $\mathbf{G}$,
$\mathbf{r}_{\alpha}=\mathbf{r}-\mathbf{R}_{\alpha}$ is the distance
from the nucleus $\alpha$, and $\Omega$ is the volume of the unit
cell. The radial function $u_{\alpha \ell}$ and its energy derivative
$\dot{u}_{\alpha \ell}$ are constructed by integrating a spherical
scalar-relativistic radial equation with the (spherically averaged)
effective KS potential $v_{\text{sp}}$ for a given energy parameter
$E_{\ell}$ \cite{KoellingJPC77}. Thus, the $u_{\alpha \ell}$ are
two-component functions, or four-component functions for
spin-polarized cases. The matching coefficients
$A^{\mathbf{K}}_{\alpha \ell m}$ and $B^{\mathbf{K}}_{\alpha \ell m}$
are determined by matching the value and slope of the large component
to the plane wave (it is assumed that the small component is zero at
the sphere boundary as the plane waves are non-relativistic (one- or
two-component) functions).

The addition of the energy derivative $\dot{u}_{\alpha \ell}$
linearizes the radial basis with respect to the energy. This
distinguishes LAPW from Slater's original APW method
\cite{slaterWaveFunctionsPeriodic1937}. It is important to note for
the discussion below that the contribution of
$B^{\mathbf{K}}_{\alpha \ell m}\dot{u}_{\alpha \ell}$ to the wave
functions (and electron density) is a measure of the quality of the
basis set. It should be much smaller than the contribution from
$A^{\mathbf{K}}_{\alpha \ell m} {u}_{\alpha \ell}$; otherwise it is an
indication that the energy parameter $E_{\alpha\ell}$ or the atomic
radius $R_{\mathrm{at}}^{\alpha}$ may be badly chosen.

An important addition to the basis set are local orbitals (LO)
\cite{SinghPRB91a}; these are linear combinations of LAPW radial
functions with a third radial function
$C^{i}_{\alpha \ell m} u_{\alpha \ell} \left( r_{\alpha},
  E^{\alpha}_{\ell i} \right)$ with another (often semi-core) energy
parameter $E^{\alpha}_{\ell i}$. Provided the energy parameters are
properly chosen (far enough apart), multiple LOs, indexed $i$, could
be added per atom and per azimuthal quantum number $\ell$.  They are
defined to be zero in the interstitial region, and the coefficients
$A^{i}_{\alpha \ell m}$, $B^{i}_{\alpha \ell m}$, and
$C^{i}_{\alpha \ell m}$ are chosen such that the linear combination is
normalized and has zero value and slope at the atomic sphere boundary.

Another linearization scheme is the APW+lo basis. It uses the LO
concept to eliminate the explicit energy-dependence of the APW basis
functions. For a detailed discussion of this approach, see Refs.
\cite{SjostedtSSC00,MadsenPRB01}.

However, all approaches boil down to linear combinations of radial
functions. For generality and simplicity of the notation, in the
following we will write basis functions using $\phi_{\mu}$ to indicate
either an extended (L)APW or a local orbital (LO/lo) basis function,
and the associated (linear combination of) radial functions simply as
$f_{\mu \ell m}$; $\mu$ is a shorthand index for, as applicable, the
atom index $\alpha$, lo index $i$, and wave number
$\mathbf{K}$. Unless necessary, we will also mute the spin index
$\sigma$ (however, all equations are given in the spin-polarized
form), the atomic index $\alpha$ as well as the $r$ and $\vu{r}$
dependencies of functions in most equations.

\subsection{Meta-generalized gradient approximation}
\label{sec:MGGA}

As mentioned in Sec.~\ref{sec:introduction}, KED-dependent MGGA functionals
\begin{equation}
\label{eq:Exc}
E_{xc} = \int\limits_{\text{cell}}\epsilon_{xc} \left(
\rho_{\uparrow}, \rho_{\downarrow} ,
\nabla \rho_{\uparrow}, \nabla \rho_{\downarrow},
{\tau}_{\uparrow} , {\tau}_{\downarrow} \right)\dd^{3}\mathbf{r}
\end{equation}
depend on the electron density
\begin{equation}
\rho_{\sigma}=\underbrace{\sum\limits_{n\mathbf{k}}w_{n\mathbf{k}}^{\sigma}\left\vert{\psi}_{n\mathbf{k}}^{\sigma}\right\vert^{2}}_{\rho_{\sigma}^{\text{val}}}+\underbrace{\sum_{\alpha n \kappa m}
\left\vert\psi_{n \kappa m}^{\alpha\sigma}\right\vert^{2}}_{\rho_{\sigma}^{\text{core}}},
\label{eq:rho}
\end{equation}
its first derivative $\nabla\rho_{\sigma}$, and the non-interacting
positive-definite KED
\begin{align}
\tau_{\sigma} = & \underbrace{\frac{1}{2}\sum\limits_{n\mathbf{k}}w_{n\mathbf{k}}^{\sigma} \nabla{\psi}_{n\mathbf{k}}^{\sigma*}\cdot\nabla{\psi}_{n\mathbf{k}}^{\sigma}}_{\tau_{\sigma}^{\text{val}}} \nonumber\\
& + \underbrace{\frac{1}{2}\sum\limits_{\alpha n\kappa m} \nabla\psi_{n \kappa m}^{\alpha\sigma*}\cdot\nabla\psi_{n \kappa m}^{\alpha\sigma}}_{\tau_{\sigma}^{\text{core}}},
\label{eq:tau}
\end{align}
where $w_{n\mathbf{k}}^{\sigma}$ is the product of the
$\mathbf{k}$-point weight and occupation number. Note that this
positive-definite KED is different from the KED that is derived from
the sum of eigenvalues in all-electron methods
\cite{YePRB15,WeinertPRB82}.

Since Eq.~(\ref{eq:Exc}) depends on $\tau_{\sigma}$, it is not an
explicit functional of the density $\rho$, like LDA and GGA
functionals. Therefore, the xc potential $v_{xc,\sigma}$, which is
defined as the functional derivative of $E_{xc}$ with respect to the
density ($v_{xc,\sigma}=\delta E_{xc}/\delta\rho_{\sigma}$), can not
be calculated straightforwardly, but only via the optimized effective
potential (OEP) method \cite{SharpPR53}. The OEP equations are often
solved approximately by using the Krieger-Li-Iafrate simplification
\cite{KriegerPRA92a} of the OEP method (see
Refs.~\cite{ArbuznikovCPL03,EichJCP14,YangPRB16}). Therefore, in most
implementations the corresponding potential of Eq.~(\ref{eq:Exc}) is
calculated within the generalized KS \cite{SeidlPRB96} (gKS) framework
by taking the functional derivative with respect to the KS orbital
${\psi}_{i}^{\sigma}$ (Ref.~\cite{NeumannMP96}):
\begin{align}
\hat{v}_{xc,\sigma}\psi_{i}^{\sigma} = &
\frac{\delta E_{xc}}{\delta\psi_{i}^{{\sigma}*}} \nonumber \\
= & \left(\frac{\partial\epsilon_{xc}}{\partial\rho_{\sigma}} -
\nabla\cdot\frac{\partial\epsilon_{xc}}{\partial\nabla\rho_{\sigma}}
\right)\psi_{i}^{\sigma} \nonumber \\
& -\frac{1}{2}\nabla\cdot\left(\frac{\partial\epsilon_{xc}}{\partial \tau_{\sigma}}
\nabla\psi_{i}^{\sigma}\right) .
\label{eq:vxc}
\end{align}
In Eq.~(\ref{eq:vxc}), the term in the large parentheses,
$v_{xc,\sigma}^{\text{mult}}=\partial\epsilon_{xc}/\partial\rho_{\sigma}-\nabla\cdot\left(\partial\epsilon_{xc}/\partial\nabla\rho_{\sigma}\right)$,
has the same form as a GGA potential
$v_{xc,\sigma}^{\text{GGA}}=\partial\epsilon_{xc}^{\text{GGA}}/\partial\rho_{\sigma}-\nabla\cdot\left(\partial\epsilon_{xc}^{\text{GGA}}/\partial\nabla\rho_{\sigma}\right)$
and is multiplicative. The expanded formula for this term can be found
in appendix~\ref{sec:mult-part-mgga}. The last term in
Eq.~(\ref{eq:vxc}) arises due to the KED-dependency of
Eq.~(\ref{eq:Exc}) and consists of a non-multiplicative operator
$\hat{v}_{\tau, \sigma}=\left(-1/2\right)\nabla\cdot\left(v_{\eta,
    \sigma}\nabla\right)$ where for conciseness we define
$v_{\eta, \sigma} = \partial\epsilon_{xc}/\partial \tau_{\sigma}$.

\subsection{Hamiltonian matrix element}
\label{sec:matrix}

It is in principle straightforward to calculate the Hamiltonian matrix
elements in a gKS approach, namely by applying the operator
$\hat{v}_{\tau}$ on a basis function $\phi_{\nu}$ and multiplying on
the left with another basis function $\phi_{\mu}$. In the LAPW method,
however, the basis functions themselves depend on the potential and
the gKS approach cannot be applied to the construction of the radial
basis functions (nor to the core electrons) because they are
calculated by direct integration. If a radial equation
  (relativistic or not) is integrated using a non-multiplicative
  potential, the solutions are in general not orthogonal. Therefore,
we use an appropriate GGA xc potential for this step.

We note that it may be possible to use an OEP (or an approximation
thereof) during this step, while keeping the gKS scheme to compute
the matrix elements. This lies outside the scope of this paper,
however.

This approach has some limitations. First, if the GGA potential is not
well-chosen, the quality of the (radial) basis functions is
diminished. In practice (as will be shown in
Sec.~\ref{sec:validation}) this is rarely an issue. A previous work
has already shown that the OEP $v_{xc}$ of SCAN only differs in small
details from the potential $v_{xc}$ of PBE \cite{YangPRB16}. If
problems do occur due to poor basis functions, this can be
detected. In this case (which can occur also in typical KS
calculations, for example when energy parameters are badly chosen) the
charge contribution coming from the linearizing term
$\dot{u}_{\alpha \ell}$ becomes large. Our code then produces a
warning automatically. Examples of this will be discussed in
Sec.~\ref{sec:validation}.

Secondly, because the core and valence electrons are
treated using inconsistent potentials (which are not the functional
derivative of a single energy functional) the calculation of forces is
in principle not possible. Note that the valence electrons are still
treated fully consistently with the MGGA potential in the gKS scheme.

The GGA xc potential for the core electrons and radial functions
$f_{\mu\ell m}$ was chosen according to the variational
principle. In Ref.~\cite{TranJCP19}, the band gap of solids was
calculated non-self-consistently (from total-energy calculations) with
a MGGA, but using GGA orbitals. It was shown that the variationally
optimal GGA orbitals (those giving the lowest total MGGA energy) are
also the ones that lead to the most correct MGGA band gaps, which is
rather expected. For the present work, it was found that the best GGA
potential to use in the construction of the basis functions is the
same as those that were determined in Ref.~\cite{TranJCP19} to be the
best for non-self-consistent MGGA calculations (and in the subsequent work
\cite{TranPRB20}), namely, RPBE \cite{HammerPRB99} (for TPSS
\cite{TaoPRL03} and SCAN \cite{SunPRL15}), mRPBE \cite{TranJCP19} (for
HLE17 \cite{VermaJPCC17}), and HCTH/407 \cite{BoeseJCP01} (for TASK
\cite{AschebrockPRR19}).

Note that using only the multiplicative part of the MGGA potential in
this step is not possible. The contribution from the non-multiplicative
part to the total potential in Eq.~(\ref{eq:vxc}) is so large that the remaining
multiplicative part is a very poor approximation, much worse than a
standard GGA potential. Ignoring the non-multiplicative part in this
step introduces large errors in the core density and the radial
functions.

With the basis set fully determined, we can set up the secular equation to
solve the gKS eigenvalue problem
$\hat{H} \psi_n = \epsilon_n \psi_{n}$. The overlap matrix $S$ is
unaffected by the choice of the potential. It is natural in APW based
methods to consider the spherical part of the Hamiltonian separately
from the non-spherical part and the interstitial part. For a MGGA
potential in the gKS scheme, a third contribution is given by the
matrix elements of the KED-derived operator $\hat{v}_{\tau}$:
\begin{equation}
  \label{eq:hamiltionan} \hat{H} = \hat{H}_{I} + \hat{H}_{sp} + v_{ns} + \hat{v}_{ns, \tau}.
\end{equation}
In APW based methods the contribution of the spherical part
$\hat{H}_{sp}$ to the matrix elements is fully determined by the
energy parameters $E_{l}$ and orthonormality of the radial
functions. The (atomic) non-spherical and KED-dependent contributions
are calculated through numerical integration. The first two terms are
discussed in the literature \cite{Singh}, we only note that a
correction must be added to the spherical part, to account for the
difference between the GGA used to construct the radial functions and
the multiplicative part of the MGGA:
\begin{align}
\label{eq:correction-sp}
  \mel**{\phi_{\mu}}{\hat{H}_{sp}}{\phi_{\nu}} = &
\mel**{\phi_{\mu}}{\hat{H}_{sp}^{\text{GGA}}}{\phi_{\nu}} \nonumber\\
& + \mel**{\phi_{\mu}}{v_{sp}^{\text{mult}} - v_{sp}^{\text{GGA}}}{\phi_{\nu}},
\end{align}
where $v_{sp}^{\text{mult}}$ and $v_{sp}^{\text{GGA}}$ are the spherical
components of the multiplicative part of the potential and the
auxiliary GGA potential, respectively. The first term of the RHS of
Eq.~(\ref{eq:correction-sp}) is still only dependent on the energy
parameters, whereas the second term is calculated by numerical
integration.

The KED-derived gKS contribution
$\mel**{\phi_{\mu}}{\hat{v}_{\tau}}{\phi_{\nu}}$ is evaluated using
integration by parts:
\begin{equation}
  \begin{split} \mel**{\phi_{\mu}}{\hat{v}_{\tau}}{\phi_{\nu}} = &
\frac{1}{2} \left[ \sum\limits_{\alpha} \int\limits_{S_{\alpha}} +
\int\limits_{\mathrm{I}} \right] v_{\eta}\nabla{\phi}_{\mu}^{*} \cdot
\nabla\phi_{\nu}\dd^{3}\mathbf{r} {} \\ & \quad - \frac{1}{2}
\sum\limits_{\alpha} \oint\limits_{\partial S_{\alpha}} v_{\eta}
\phi_{\mu}^{*} \left( \nabla \phi_{\nu}  \right) \cdot \vu{r} \dd \Omega
\end{split}
\end{equation}
and their detailed form for APW based basis functions can be found in
appendix~\ref{sec:derivation}.

With the Hamiltonian and overlap
matrices determined, the secular equation can be solved. It provides
the KS orbitals, from which the electron density and the KED are
determined, closing the self-consistency loop.

\subsection{Total energy}
\label{sec:totalenergy}

The total energy per unit cell is given by
\begin{align}
\label{eq:Etot1}
  \begin{split} E_{\text{tot}} = T_{\text{s}} & +
\frac{1}{2}\int\limits_{\text{cell}}
v_{\text{Coul}}(\mathbf{r})\rho(\mathbf{r})\dd^{3}\mathbf{r} {} \\ & -
\frac{1}{2}\sum_{\alpha}Z_{\alpha}
v_{\text{M}}^{\alpha}(\mathbf{R}_{\alpha}) + E_{xc}
\end{split}
\end{align} where $T_{\text{s}}$ is the kinetic energy of the
electrons and
\begin{align}
\label{eq:vcoul} v_{\text{Coul}}(\mathbf{r}) & =
\int\limits_{\text{cell}}\frac{\rho(\mathbf{r}')}
{\left\vert\mathbf{r}-\mathbf{r}'\right\vert}\dd^{3}\mathbf{r}' -
\sum_{\beta}
\frac{Z_{\beta}}{\left\vert\mathbf{r}-\mathbf{R}_{\beta}\right\vert},
\\
\label{eq:vM} v_{\text{M}}^{\alpha}(\mathbf{R}_{\alpha}) & =
\int\limits_{\text{cell}}\frac{\rho(\mathbf{r}')}
{\left\vert\mathbf{R}_{\alpha}-\mathbf{r}'\right\vert}\dd^{3}\mathbf{r}'
- \sum_{\beta\neq\alpha}
\frac{Z_{\beta}}{\left\vert\mathbf{R}_{\alpha}-\mathbf{R}_{\beta}\right\vert},
\end{align} are the Coulomb and Madelung potentials, respectively
($Z_{\alpha}$ is the charge of nucleus $\alpha$). By using the sum of
the eigenvalues, $E_{\text{tot}}$ can be rewritten for a MGGA
functional in the gKS scheme as
\begin{equation}
\label{eq:Etot}
\begin{split} E_{\text{tot}} = & \sum_{\alpha\sigma n\kappa m}
\epsilon_{n\kappa m}^{\alpha\sigma} + \sum_{\sigma
n\mathbf{k}}w_{n\mathbf{k}}^{\sigma}\epsilon_{n\mathbf{k}}^{\sigma} +
E_{xc}^{\text{MGGA}} \\ & -\frac{1}{2}\int\limits_{\text{cell}}
v_{\text{Coul}}(\mathbf{r})\rho(\mathbf{r})\dd^{3}\mathbf{r} -
\frac{1}{2}\sum_{\alpha}^{\text{cell}}Z_{\alpha}
v_{\text{M}}^{\alpha}(\mathbf{R}_{\alpha}) \\ &
-\sum_{\sigma}\int\limits_{\text{cell}}
v_{xc,\sigma}^{\text{GGA}}(\mathbf{r})
\rho_{\sigma}^{\text{core}}(\mathbf{r})\dd^{3}\mathbf{r} \\ &
-\sum_{\sigma}\int\limits_{\text{cell}}
v_{xc,\sigma}^{\text{mult}}(\mathbf{r})
\rho_{\sigma}^{\text{val}}(\mathbf{r})\dd^{3}\mathbf{r} \\ & -
\sum_{\sigma}\int\limits_{\text{cell}} v_{\eta, \sigma}(\mathbf{r})
\tau_{\sigma}^{\text{val}}(\mathbf{r})\dd^{3}\mathbf{r},
\end{split}
\end{equation}
where $v_{xc,\sigma}^{\text{GGA}}$ is the GGA potential used for the
core electrons (as well as for the calculation of the radial functions
$f_{\mu\ell m}$). Except for the addition of the last KED-derived term,
and the separation of the core and valence potentials, this expression
is equivalent to the one given by \citet{WeinertPRB82} for the KS scheme.

\subsection{Computational details}
\label{sec:computationaldetails}

As already mentioned, the WIEN2k code \cite{WIEN2k,BlahaJCP20} is used
for the present study. In the self-consistent implementation of MGGA
functionals, the xc energy density $\epsilon_{xc}$ and its derivatives
with respect to $\rho_{\sigma}$ and $\tau_{\sigma}$ are provided by
the library of xc functionals Libxc \cite{MarquesCPC12,LehtolaSX18}.

\begin{table}[ht]
  \caption{\label{tab:parameters}Parameters required to converge total
    energy to 0.01~Ry, band gaps to 0.01~eV, EFG to
    0.01~$10^{21}$~V/m$^{2}$, and the magnetic moment to
    0.01~$\mu_B$. The column names $G$ and $K$ are short for the
    parameters $G_{\mathrm{max}}$ and
    $R^{\mathrm{at}}_{\mathrm{min}}K_{\mathrm{max}}$ as defined in the text.}
  \begin{center}
    \begin{tabular*}{\linewidth}{@{\extracolsep{\fill}}lrrrrrrrrrr}
    \hline\hline
      & \multicolumn{2}{c}{PBE} & \multicolumn{2}{c}{TPSS} & \multicolumn{2}{c}{SCAN}
      & \multicolumn{2}{c}{TASK} & \multicolumn{2}{c}{HLE17} \\
      Solid & $G$ & $K$ & $G$ & $K$ & $G$ & $K$ & $G$ & $K$ & $G$ & $K$ \\
\hline
\ch{Al2O3} & 12 & 8.0 & 12 & 8.0 & 20 & 8.0 & 20 & 8.5 & 12 & 8.0\\
LiF & 12 & 6.0 & 16 & 7.0 & 16 & 6.0 & 20 & 8.0 & 16 & 7.0\\
Ru & 12 & 9.0 & 12 & 8.5 & 16 & 9.0 & 16 & 9.0 & 12 & 9.0\\
Si & 12 & 7.0 & 12 & 7.0 & 12 & 7.0 & 12 & 7.0 & 12 & 7.0\\
Zn & 12 & 8.5 & 12 & 9.0 & 24 & 9.0 & 12 & 9.0 & 12 & 9.0\\
NiO & 12 & 8.0 & 12 & 8.0 & 22 & 8.0 & 12 & 7.5 & 12 & 8.0\\
\ch{MoS2} & 10 & 7.0 & 10 & 7.0 & 10 & 7.0 & 10 & 7.0 & 14 & 7.0\\
      Ge & 12 & 8.0 & 12 & 8.0 & 12 & 8.0 & 12 & 8.0 & 12 & 8.0\\
      \hline \hline
    \end{tabular*}
  \end{center}
\end{table}

An overview of required parameters to reach the same convergence
level for PBE, SCAN, HLE17, and TASK are listed in
Table~\ref{tab:parameters}. $G_{\mathrm{max}}$ is the cutoff for
the fourier coefficients of quantities (density, KED, and potentials)
in the interstitial (and thus the size of the fourier grids), whereas
$R^{\mathrm{at}}_{\mathrm{min}}K_{\mathrm{max}}$ is the product of the
plane wave basis cutoff and the smallest atomic sphere in the
system. Other parameters like the {$\mathbf{k}$-mesh}, and cutoffs of
the spherical harmonics expansions were found to be unaffected by the
choice for the choice between a MGGA or a GGA functional.

It is clear that SCAN and TASK require a larger fourier cutoff
$G_{\mathrm{max}}$. This applies to a lesser extent to HLE17. The
required plane wave cutoff are not significantly affected (though
there are some fluctuations), except for LiF. LiF is a special case
however, as it has no core states, and a convergence of 0.01~eV in the
band gap (considering the predicted band gap of 12.58~eV with TASK) is a
quite strict convergence criterion.

On the basis of these results, a $G_{\mathrm{max}}$ of 24 should be
safe for all or most MGGAs, compared to 12 for PBE (although 14 is the
default setting in WIEN2k). SCAN is known to be numerically demanding
and regularized versions are available
\cite{furnessAccurateNumericallyEfficient2020a,bartokRegularizedSCANFunctional2019}
which should have better convergence characteristics.  Depending on
the MGGA functional that is chosen and the system under consideration,
we expect that in most cases a lower value can be used for
$G_{\mathrm{max}}$. The choice for basis set size (which is the most
important parameter w.r.t. computational time as it determines the
size of the Hamiltonian matrix) is not affected by MGGAs compared to
PBE.

\begin{table}[ht]
  \caption{\label{tab:timing}Timings per iteration (in seconds), and
    number of iterations needed from an initial superposition of
    atomic densities, for four benchmark systems using PBE, SCAN, and
    HLE17. We performed all calculations sequentially (i.e. on a
    single core) on an Intel i7-7820X CPU (8 cores @ 3.60 GHz), except
    the Ge (d) case which was parallelized using MPI across all cores
    of 2 identical
    machines. $R^{\mathrm{at}}_{\mathrm{min}}K_{\mathrm{max}}$ was
    chosen the same for all functionals and according to the values
    listed in Table~\ref{tab:parameters}: $7.0$ for Si and \ch{MoS2},
    $8.0$ for Ge and \ch{Al2O3}, and $9.0$ for Zn. The timings for one
    iteration have been separated into the calculation of the
    potential $V$, the set-up and diagonalization of the Hamiltonian
    matrix, and `other' (calculation of density, KED, mixing, and core
    states). For \ch{MoS2}, spin-orbit coupling was included, which
    explains the larger amount of time spent in `other'. The $^{*}$
    for the number of iterations for \ch{MoS2} using TASK signifies
    that this case was restarted from a converged SCAN calculation.}
  \centering
  \begin{tabular*}{\linewidth}{@{\extracolsep{\fill}}llrrrrr}
\hline \hline
 Functional & step  & Si & Zn & \ch{Al2O3} & \ch{MoS2} & Ge (d)\\
\hline
 & $V$ & 1 & $<1$ & 1 & 2 & 44\\
{PBE} & $H$ & 2 & 7 & 172 & 38 & 882\\
{($G_{\mathrm{max}}=12$)} & other & 2 & 4 & 57 & 87 & 287\\
 & \# iter. & 9 & 9 & 8 & 11 & 13\\
\hline
 & $V$ & 7 & 3 & 6 & 8 & 334\\
 {PBE}& $H$ & 2 & 7 & 176 & 38 & 874\\
{($G_{\mathrm{max}}=24$)} & other & 2 & 4 & 57 & 85 & 295\\
 & \# iter. & 9 & 9 & 8 & 11 & 13\\
\hline
  & $V$ & 15 & 4 & 14 & 25 & 605\\
 {SCAN}& $H$ & 2 & 7 & 227 & 41 & 885\\
 {($G_{\mathrm{max}}=24$)}& other & 2 & 4 & 56 & 82 & 297\\
 & \# iter. & 9 & 10 & 12 & 11 & 15\\
\hline
{TPSS} & $V$ & 19 & 6 & 17 & 23 & 626\\
 {($G_{\mathrm{max}}=24$)}& \# iter. & 9 & 9 & 10 & 11 & 14\\
\hline
{TASK} & $V$ & 13 & 4 & 11 & 18 & 581\\
 {($G_{\mathrm{max}}=24$)}& \# iter. & 9 & 11 & 11 & $65^{*}$ & 15\\
\hline
{HLE17} & $V$ & 22 & 6 & 19 & 26 & 634\\
 {($G_{\mathrm{max}}=24$)}& \# iter. & 9 & 11 & 11 & 11 & 17\\
\hline
  \end{tabular*}
\end{table}

In Table~\ref{tab:timing}, we list the timings for five systems. Three
of them, \ch{Al2O3}, Si, and Zn will be discussed below. The other two
systems have been chosen to include also more computationally
demanding cases. \ch{MoS2} is a well-known transition metal monolayer,
and calculations including MGGAs were published in
\cite{TranJCP21}. The last system is a supercell of Ge with an
interstitial defect. It contains 129 atoms in the unit cell, and has
been discussed in~\cite{murphy-armandoLightEmissionDirect2021} (not
including MGGA calculations).

We show timings for PBE, PBE using the recommended MGGA parameters,
and SCAN. This way, the influence of the parameters can be separated
from the computational complexity that is intrinsic to the MGGAs. For
TPSS, TASK, and HLE17 we show just the timing for the construction of
the potential and number of self-consistency iterations required, as
the other timings are indistinguishable.

The largest influence is obviously on the calculation of the
potential. The doubling of $G_{\mathrm{max}}$ implies that the Fourier
grid will be eight times larger. Additionally, a second Fourier grid
must be stored in memory for the KED dependent part of the
potential. This is reflected clearly in the timings, where one sees a
large jump when doubling $G_{\mathrm{max}}$ using PBE. The additional
complexity introduced by using a MGGA is a factor of 2--3, depending
on the chosen functional. In general, HLE17 takes the longest to
evaluate, followed by TPSS, SCAN, and finally TASK. The lower relative
complexity of TASK might be explained by the fact that only the
exchange part is a MGGA (the correlation is taken from LDA). It is
interesting to note that for \ch{MoS2} the relative order of SCAN and
TPSS is reversed. This is probably caused by the presence of vacuum
regions in this case, combined with the piecewise definitions of the
functionals. It is hard to track down the exact cause however, due to
the complexity of the definitions of the functionals and the impact of
compiler optimizations. Overall the main impact comes from the
parameter $G_{\mathrm{max}}$, for which we chose for the timings a
very high value (compare with the actual necessary $G_{\mathrm{max}}$
as given in Table~\ref{tab:parameters}). For very intensive cases, it
will often be possible to reduce this parameter depending on the
composition of the system and functional used, e.g. after testing the
convergence on a smaller system with similar composition or physical
features.

The set-up and diagonalization are only modestly affected. In fact, the
diagonalization should be exactly as expensive, as the Hamiltonian
matrix has the same dimensions. The change in this step comes almost
entirely from the calculation of the spherical terms (see
Eq.~(\ref{eq:mel-volume})), which takes about 50\,\% longer. Because
the set-up has a computational complexity of
$\mathcal{O}\left( N^{2} \right)$, but for the diagonalization it is
$\mathcal{O}\left( N^3 \right)$, the difference between the timings
for this step will tend to zero for larger cases.

In the last category `other', there are very small differences, which
can be traced back to the construction and mixing of the
positive-definite KED $\tau$ (which we implemented similarly to
\cite{YePRB15}).

The number of iterations needed was unchanged, or only
modestly increased. The exception was the calculation of \ch{MoS2}
with TASK, where 65 iterations were needed starting from a converged
PBE calculation (compared to the others, which are started from a
superposition of atomic densities). This shows that, in some cases,
convergence with gKS MGGAs can be tricky for 2D systems (and
probably also other systems including vacuum regions). The authors
also experienced this for the calculation presented in
Ref.~\cite{TranJCP21}. However, we did not observe such
issues in bulk systems.

On the whole, we see that the computational expense of a MGGA
calculation using our implementation is not much higher than that of a
GGA calculation. On a few points, there are probably still
optimizations possible. Notably, the mixing of the density in our code
is done using a sophisticated optimization
scheme~\cite{marksPredictiveMixingDensity2021}, whereas the KED is
mixed independently according to a simple `Pratt' scheme, where the
current and previous KED are simply added with a fixed non-adaptive
proportion and scaled. Possibly an improved mixing scheme could solve
the slow convergence found for some cases with vacuum regions.

\section{Validation of the implementation}
\label{sec:validation}

\subsection{Band gap}
\label{sec:bandgap}

\begin{table*}
  \caption{\label{tab:bandgap1}Band gaps (in eV) calculated with MGGA
    functionals using the WIEN2k code at experimental lattice
    parameters. The columns $\Delta$ show the difference with respect
    to the VASP result
    $\left(E_{g}^{\textsc{WIEN2k}}-E_{g}^{\textsc{VASP}}\right)$ from
    Refs.~\cite{BorlidoJCTC19,BorlidoNPJCM20} (except AlSb, which we
    calculated ourselves, see text). For comparison, the PBE results
    are also shown.}
  \begin{tabular*}{\linewidth}{@{\extracolsep{\fill}}lSSSSSSSSSS}
    \hline\hline
    {Solid} & {PBE} & {$\Delta$} & {TPSS} & {$\Delta$} & {SCAN} & {$\Delta$} & {HLE17} & {$\Delta$} & {TASK} & {$\Delta$} \\
  \hline
  \ch{Al2O3} & 6.20 & 0.01 & 6.35 & 0.04 & 7.08 & 0.04 & 7.12 & 0.17 & 8.70 & -0.01 \\
  \ch{AlAs} & 1.47 & 0.04 & 1.53 & 0.04 & 1.77 & 0.04 & 2.53 & 0.04 & 2.49 & 0.05 \\
  \ch{AlN} & 4.14 & -0.00 & 4.14 & 0.00 & 4.79 & 0.00 & 4.79 & 0.05 & 5.77 & -0.08 \\
  \ch{AlP} & 1.59 & 0.01 & 1.66 & 0.02 & 1.91 & 0.01 & 2.73 & -0.06 & 2.45 & -0.01 \\
  \ch{AlSb} & 1.22 & 0.01 & 1.28 & 0.01 & 1.37 & 0.00 & 1.94 & 0.14 & 2.15 & 0.08 \\
  \ch{Ar} & 8.71 & -0.01 & 9.37 & 0.06 & 9.63 & 0.13 & 10.90 & 0.06 & 13.21 & -0.06 \\
  \ch{BeO} & 7.37 & 0.02 & 7.39 & 0.06 & 8.21 & 0.05 & 8.51 & 0.04 & 9.59 & -0.04 \\
  \ch{BN} & 4.46 & 0.01 & 4.49 & 0.06 & 4.96 & 0.03 & 5.73 & 0.04 & 5.42 & 0.02 \\
  \ch{BP} & 1.25 & -0.02 & 1.29 & -0.02 & 1.53 & -0.05 & 2.16 & -0.07 & 1.48 & -0.02 \\
  \ch{C} & 4.14 & -0.01 & 4.17 & -0.02 & 4.54 & -0.03 & 5.01 & 0.01 & 4.33 & -0.00 \\
  \ch{CaF2} & 7.28 & 0.00 & 7.76 & 0.03 & 7.88 & 0.05 & 9.38 & 0.04 & 10.45 & 0.03 \\
  \ch{CaO} & 3.67 & 0.04 & 3.79 & 0.04 & 4.24 & 0.08 & 4.54 & 0.03 & 5.16 & -0.08 \\
  \ch{CdSe} & 0.71 & -0.04 & 0.90 & -0.00 & 1.11 & 0.04 & 1.67 & -0.06 & 2.11 & 0.00 \\
  \ch{GaAs} & 0.52 & -0.05 & 0.68 & -0.00 & 0.78 & -0.02 & 0.75 & 0.09 & 1.72 & 0.05 \\
  \ch{GaP} & 1.59 & -0.05 & 1.64 & -0.04 & 1.83 & -0.05 & 2.22 & 0.02 & 2.37 & -0.01 \\
  \ch{Ge} & 0.06 & -0.05 & 0.18 & 0.01 & 0.18 & 0.04 & 0.00 & 0.00 & 0.89 & 0.02 \\
  \ch{InP} & 0.68 & -0.03 & 0.82 & 0.00 & 1.04 & -0.01 & 1.10 & -0.02 & 1.87 & -0.21 \\
  \ch{KCl} & 5.21 & -0.00 & 5.73 & 0.01 & 5.84 & 0.06 & 6.91 & -0.01 & 8.72 & -0.04 \\
  \ch{Kr} & 7.26 & -0.01 & 7.89 & -0.00 & 8.03 & -0.01 & 9.30 & -0.01 & 11.37 & -0.11 \\
  \ch{LiCl} & 6.33 & 0.00 & 6.57 & 0.02 & 7.28 & 0.10 & 7.79 & 0.02 & 9.50 & 0.04 \\
  \ch{LiF} & 9.08 & 0.00 & 9.25 & -0.01 & 9.97 & -0.01 & 10.81 & -0.01 & 12.76 & 0.16 \\
  \ch{LiH} & 3.03 & 0.03 & 3.44 & 0.08 & 3.63 & -0.01 & 4.75 & 0.13 & 5.48 & 0.05 \\
  \ch{MgO} & 4.71 & 0.00 & 4.83 & 0.03 & 5.58 & 0.05 & 5.70 & 0.09 & 7.25 & -0.06 \\
  \ch{NaCl} & 5.11 & 0.01 & 5.49 & 0.04 & 5.89 & 0.04 & 6.78 & 0.07 & 8.66 & -0.04 \\
  \ch{NaF} & 6.33 & 0.02 & 6.75 & 0.07 & 7.04 & 0.02 & 8.41 & 0.12 & 10.32 & -0.04 \\
  \ch{Ne} & 11.58 & -0.00 & 12.17 & 0.02 & 12.90 & 0.13 & 14.29 & 0.03 & 16.98 & -0.01 \\
  \ch{Si} & 0.58 & -0.04 & 0.66 & -0.04 & 0.83 & -0.04 & 1.56 & -0.07 & 1.00 & -0.02 \\
  \ch{SiC} & 1.36 & 0.01 & 1.38 & 0.06 & 1.68 & -0.03 & 2.34 & 0.05 & 1.95 & -0.03 \\
  \ch{ZnO} & 0.82 & 0.02 & 0.76 & 0.03 & 1.18 & 0.04 & 2.22 & -0.16 & 2.04 & -0.04 \\
    \ch{ZnS} & 2.12 & -0.03 & 2.28 & -0.01 & 2.63 & 0.00 & 3.18 & -0.11 & 3.77 & -0.02 \\
    \hline
  MD & 0.00 & & -0.02 & & -0.02 & & -0.02 & & 0.04 \\
  MAD & 0.02 & & 0.03 & & 0.04 & & 0.06 & & 0.07 \\
    MAPD & 4.2\,\% & & 1.4\,\% & & 2.1\,\% & & 2.1\,\% & & 2.4\,\% \\
    \hline \hline
\end{tabular*}
\end{table*}

\begin{table*}
  \caption{\label{tab:bandgap2}Influence on MGGA band gaps of the
    chosen GGA potential (PBE or the optimal one) to calculate the
    core electrons and radial functions $f_{\mu\ell m}$. The optimal
    potential is RPBE for TPSS and SCAN, mRPBE for HLE17, and HCTH/407
    for TASK. A positive value means that the band gap with the
    optimal GGA potential is larger. For entries with a star $^{*}$,
    we needed to add additional basis functions (in the form of HDLOs)
    to properly converge the calculations (see text and
    Sec.~\ref{sec:APW} for details). The values are in eV.}
\begin{tabular*}{\linewidth}{@{\extracolsep{\fill}}lSSSSSS}
  \hline \hline
    {Solid} & {TPSS(PBE-RPBE)} & {SCAN(PBE-RPBE)} & {HLE17(PBE-mRPBE)}
      & {TASK(PBE-HCTH/407)} \\
    \hline
    \ch{Al2O3} & -0.01 & -0.01 & 0.01 & 0.00 \\
    \ch{AlAs} & -0.00 & -0.00 & 0.05 & -0.02 \\
    \ch{AlN} & -0.00 & -0.00 & -0.03 & 0.00 \\
    \ch{AlP} & -0.00 & -0.00 & 0.08 & 0.00 \\
    \ch{AlSb} & -0.00 & 0.00 & -0.02 & -0.01 \\
    \ch{Ar} & -0.01 & -0.01 & 0.15 & -0.00 \\
    \ch{BeO} & -0.00 & -0.01 & 0.01$^{*}$ & -0.00 \\
    \ch{BN} & -0.00 & -0.00 & 0.03 & -0.01 \\
    \ch{BP} & -0.00 & -0.00 & 0.07 & 0.01 \\
    \ch{C} & -0.00 & -0.00 & 0.01 & -0.00 \\
    \ch{CaF2} & -0.02 & -0.02 & 0.14$^{*}$ & 0.13 \\
    \ch{CaO} & -0.01 & -0.01 & 0.06$^{*}$ & 0.03 \\
    \ch{CdSe} & -0.00 & -0.00 & 0.05$^{*}$ & 0.01 \\
    \ch{GaAs} & -0.00 & -0.01 & 0.02$^{*}$ & 0.01 \\
    \ch{GaP} & -0.00 & -0.00 & 0.04$^{*}$ & 0.01 \\
    \ch{Ge} & -0.01 & -0.00 & 0.00 & -0.01 \\
    \ch{InP} & -0.01 & -0.00 & -0.02$^{*}$ & -0.00 \\
    \ch{KCl} & -0.01 & -0.01 & 0.01$^{*}$ & 0.02 \\
    \ch{Kr} & -0.01 & -0.01 & 0.12 & 0.05 \\
    \ch{LiCl} & -0.01 & -0.01 & 0.06 & 0.02 \\
    \ch{LiF} & -0.02 & -0.02 & 0.02$^{*}$ & 0.05 \\
    \ch{LiH} & 0.00 & 0.00 & -0.00 & -0.00 \\
    \ch{MgO} & -0.01 & -0.01 & 0.02$^{*}$ & 0.02 \\
    \ch{NaCl} & -0.01 & -0.01 & 0.02$^{*}$ & 0.02 \\
    \ch{NaF} & -0.02 & -0.02 & 0.00$^{*}$ & 0.06 \\
    \ch{Ne} & -0.01 & -0.01 & 0.15$^{*}$ & 0.14 \\
    \ch{Si} & -0.00 & -0.00 & 0.09 & 0.01 \\
    \ch{SiC} & -0.00 & -0.00 & 0.06 & -0.01 \\
    \ch{ZnO} & 0.01 & 0.00 & 0.12$^{*}$ & 0.22 \\
    \ch{ZnS} & 0.00 & 0.00 & 0.17$^{*}$ & 0.11$^{*}$ \\
    \hline \hline
  \end{tabular*}
\end{table*}

In order to check the correctness of our implementation of the MGGA
potential, we consider the band gap of a set of solids. By band gap,
we mean the difference between the eigenvalues
$\epsilon_{n\mathbf{k}}$ at the valence band maximum (VBM) and
conduction band minimum (CBM):
\begin{equation}
E_{g}^{\text{(g)KS}} =
\epsilon_{\text{CBM}}^{\text{(g)KS}} -
\epsilon_{\text{VBM}}^{\text{(g)KS}},
\label{eq:EgKS}
\end{equation} which is, here for MGGA functionals, applied in the gKS
framework. At this point, it is worth mentioning a few words about the
xc derivative discontinuity $\Delta_{xc}$
\cite{PerdewPRL82,ShamPRL83}. $\Delta_{xc}$ is defined as
\begin{equation} \Delta_{xc} = E_{g}^{I-A} - E_{g}^{\text{KS}},
\end{equation} where $E_{g}^{I-A}=I-A$ is the true band gap calculated
as the difference between the ionization potential $I$ and electron
affinity $A$ of the system and $E_{g}^{\text{KS}}$ is calculated with
Eq.~(\ref{eq:EgKS}) when the potential is implemented in the KS
framework so that it is multiplicative. For solids, LDA and GGA
functionals lead to $\Delta_{xc}=0$ (see
Refs.~\cite{KraislerJCP14,GorlingPRB15}), which is the main reason why
LDA/GGA strongly underestimate the band gap with respect to experiment
\cite{PerdewIJQC86,HeydJCP05} (note, however that a few less common
GGA functionals \cite{ArmientoPRL13,FinzelIJQC17,VermaJPCL17} lead to
much better band gaps, although $\Delta_{xc}$ is still
zero). Functionals that lead to a non-multiplicative gKS potential,
like MGGAs and HF/hybrids, possess a non-zero xc derivative
discontinuity $\Delta_{xc}$ \cite{KuemmelRMP08}. With such
non-multiplicative gKS potentials, $\Delta_{xc}$ is included in
$E_{g}^{\text{gKS}}$ (see
Refs.~\cite{KuemmelRMP08,YangJCP12,PerdewPNAS17}) and, consequently,
$E_{g}^{\text{gKS}}$ is usually in better agreement with the
experimental value of $E_{g}^{I-A}$, as shown in numerous benchmark
studies \cite{SeidlPRB96,HeydJCP05,CrowleyJPCL16,GarzaJPCL16,VermaJPCL17,BorlidoJCTC19,AschebrockPRR19,BorlidoNPJCM20}.
It is also interesting to note that the derivative discontinuity $\Delta_{xc}$
is not included in the CBM-VBM difference when the MGGA
potential is implemented using the OEP method, however $\Delta_{xc}$ is in
principle non-zero and can be calculated~\cite{YangPRB16}.

Among the numerous proposed MGGA functionals \cite{DellaSalaIJQC16},
HLE17 \cite{VermaJPCC17} and TASK \cite{AschebrockPRR19} are some of
the most accurate for the band gap of solids (albeit they are slightly
less accurate than the modified Becke-Johnson (mBJ) MGGA potential
\cite{TranPRL09,KollerPRB11,TranJAP19}, as shown in
Ref.~\cite{BorlidoNPJCM20}). Therefore, they are of particular
interest for testing our implementation of MGGA potentials. Also
considered are the well-known TPSS \cite{TaoPRL03} and SCAN
\cite{SunPRL15}, the latter being very successful for total-energy
calculations \cite{ZhangNPJCM18,IsaacsPRM18}. While TPSS leads
basically to no improvement with respect to PBE \cite{PerdewPRL96} for
the band gap, SCAN is clearly more accurate, but not as much as HLE17
or TASK \cite{JanaJCP18a,BorlidoNPJCM20}.

The results for the band gap of 30 solids obtained with the WIEN2k
code are shown in Table~\ref{tab:bandgap1}. This set of solids is the
one that we used in Ref.~\cite{TranJCP19} for the non-self-consistent
calculation of the band gap with MGGA functionals using the total
energy $E_{\text{tot}}$. It is a subset of the much larger set of 473
solids built by Borlido \textit{et
  al}. \cite{BorlidoJCTC19,BorlidoJCTC20,BorlidoNPJCM20}. The results
are compared with the results from
Refs.~\cite{BorlidoJCTC19,BorlidoNPJCM20} which were obtained with the
VASP code \cite{KressePRB96} that uses the projector augmented wave
(PAW) formalism
\cite{blochlProjectorAugmentedwaveMethod1994,KressePRB99}. The
agreement between the two codes can be considered as satisfying, since
in the majority of cases the difference
$E_{g}^{\textsc{WIEN2k}}-E_{g}^{\textsc{VASP}}$ between the two codes
is below 0.05~eV. The mean absolute difference (MAD) is 0.03 and
0.04~eV for TPSS and SCAN, respectively, but slightly larger for HLE17
and TASK (0.06 and 0.07~eV, respectively). The mean differences (MD)
are for all functionals much smaller, showing there is no systematic
error. The mean absolute percentage deviation (MAPD) for the MGGA
functionals varies between 1.4 and 2.4\,\%, which is much smaller than
the one for PBE. The MAPD of 4.2\,\% of PBE however is dominated by the
very large relative difference of 88\,\% for Ge. From such small MADs
and MAPDs between WIEN2k and VASP we can conclude that the new
implementation of the MGGA potentials in WIEN2k is correct and
accurate.

For AlSb, we did not use the result from
Refs.~\cite{BorlidoJCTC19,BorlidoNPJCM20} as we found a significant
discrepancy between the results. We found a value of 2.15~eV (WIEN2k),
compared to their result of 2.87~eV (VASP). Therefore we recalculated
this case ourselves with VASP and found a value of 2.07~eV, in good
agreement with WIEN2k. Additionally we confirmed that our VASP result
can be reproduced using different pseudopotentials. Our WIEN2k result
is also independent of various parameters such as the choice of LAPW
or APW+lo basis, putting the Sb 4$s$ and 4$p$ orbitals in core or
valence, the sphere size, as well as the choice of the local GGA
potential (as indicated in Table~\ref{tab:bandgap2}).

In a recent study \cite{BorlidoJCTC20} the band gaps of the
aforementioned 473 solids were calculated with three DFT codes, namely
WIEN2k, VASP, and ABINIT \cite{GonzeCPC20}, and compared. The goal was
to estimate the error in the band gap induced by using an inconsistent
pseudopotential (PP) or PAW-setup, like for instance using a LDA or
PBE PP/PAW-setup for a calculation with another functional. The WIEN2k
results were used as reference. When a consistent PP/PAW-setup is
used, the MAD between WIEN2k and VASP/ABINIT is 0.02-0.03~eV, which is
very small. However, the MAD increases up to $\sim0.1$~eV when a LDA
or PBE PP/PAW-setup is used for a calculation with a very different
functional like HLE16 \cite{VermaJPCL17}, Sloc \cite{FinzelIJQC17}, or
mBJ. Considering this, the MAD obtained here for the MGGAs are very
reasonable, including HLE17 and TASK which are by construction very
different from more standard functionals like TPSS or SCAN.

As discussed in Sec.~\ref{sec:MGGA}, the MGGA potential is not
implemented in the atomic codes that are used to calculate the core
orbitals and the radial functions of the basis set.  Thus, a GGA
potential has to be used instead. This may lead to a suboptimal core
density and/or basis set in the spheres.  However, the variational
principle guarantees that the lowest total energy corresponds to the
best core density and basis set.  Following these principles, the
optimal GGA potential is RPBE for TPSS and SCAN, mRPBE for HLE17, and
HCTH/407 for TASK \cite{TranJCP19}.  In order to illustrate the effect
of the GGA potential used in the atomic codes on the band gap, the
calculations were repeated by using the PBE GGA potential, instead,
and Table~\ref{tab:bandgap2} shows the difference with respect to the
results in Table~\ref{tab:bandgap1}.  The effect is absolutely
negligible in the case of the TPSS and SCAN functionals. With TASK and
HLE17, differences in the range 0.1--0.2~eV are obtained for some
cases.  For ZnS with TASK and for a number of cases with HLE17 it was
necessary to improve the flexibility of the basis set by adding HDLOs
(indicated with the $^{*}$ in Table~\ref{tab:bandgap2}). For these
cases, a warning occurs because that the radial basis functions inside
the atomic spheres are inaccurate. This warning is given when the
linearizing term $\dot{u}_{\ell}$ of the basis function contributes a
larger fraction of the charge density than expected (more than a few
percent). When the radial basis functions are accurate the radial
solution $u_{\ell}$ (or linear combinations thereof) will be very
close to the KS orbitals, such that the contribution from the
linearizing term stays small. We would like to stress that these
warnings are given automatically and are a standard warning (in the
WIEN2k code) that is used to diagnose badly chosen parameters (like
atomic sphere size or energy parameters) or ghost bands.

As mentioned above, Ref.~\cite{TranJCP19} reports band gaps calculated
non-self-consistently with MGGA functionals. The orbitals (and thus
the electron density) were generated by a GGA potential and then
plugged into the total MGGA energy $E_{\text{tot}}$ to calculate the
band gap with
$E_{g}^{I-A} = I(N) - A(N) =
[E_{\text{tot}}(N-1)-E_{\text{tot}}(N)]-[E_{\text{tot}}(N)-E_{\text{tot}}(N+1)]$.
By doing so, the agreement with the self-consistent VASP results is
quite accurate provided that the optimal GGA orbitals are
used. However, this method may be more cumbersome than the
self-consistent implementation considered here. It may also be less
accurate when the orbitals found using the GGA potential differ
strongly from the self-consistent MGGA ones, which can be seen in the
variational energies. Additionally, the agreement with VASP is not as
good as when the calculations are done self-consistently. Indeed,
non-self-consistently the MAD with respect to VASP results are (in eV)
0.06 (for TPSS), 0.10 (for SCAN), 0.10 (for HLE17), and 0.20 (for
TASK). These MAD are larger than those from Table~\ref{tab:bandgap1}
that are in the range 0.03--0.07~eV.

In summary, the band gaps obtained with our self-consistent
implementation of the MGGA functionals in the WIEN2k code are in
excellent agreement with the results obtained with the VASP code. This
gives us confidence about the reliability of the implementation in
terms of correctness and accuracy.

\subsection{Magnetism}
\label{sec:magnetism}

\begin{table*}
\caption{\label{tab:moment}Spin magnetic moment $\mu_{S}$ in FM (in
$\mu_{\text{B}}$ and per formula unit) and AFM (in $\mu_{\text{B}}$
and inside the atomic volume of the transition-metal atom defined
according to the Bader volume) solids calculated with MGGA functionals
using the WIEN2k code. The columns $\Delta$ show the difference with
respect to the value from Ref.~\cite{TranPRB20} calculated without the
self-consistent MGGA potential, but with the FSM (for FM solids) or
$C$-shift (for AFM solids) method, and a negative value means that the
value from the FSM/$C$-shift method is larger. For comparison, the PBE
results are also shown.}
\begin{tabular*}{\linewidth}{@{\extracolsep{\fill}}lSSSSSSSSSS}
  \hline \hline
    {Solid} & {PBE} & {TPSS} & $\Delta$ & {SCAN} & $\Delta$
    & {HLE17}  & $\Delta$ & {TASK} & $\Delta$ \\
    \hline \bf{FM} \\
    \ch{Fe} & 2.22 & 2.23 & 0.00 & 2.60 & -0.03 & 2.67 & 0.00 & 2.76 & 0.01 \\
    \ch{Co} & 1.62 & 1.65 & 0.00 & 1.77 & -0.02 & 1.72 & 0.00 & 1.84 & 0.01 \\
    \ch{Ni} & 0.64 & 0.66 & 0.00 & 0.77 & 0.01 & 0.65 & 0.00 & 0.78 & 0.02 \\
    \bf{AFM} \\
    \ch{MnO} & 4.40 & 4.42 & 0.01 & 4.52 & -0.01 & 4.63 & 0.01 & 4.63 & 0.00 \\
    \ch{FeO}\footnotemark[1] & NA & 3.49 & NA & 3.57\footnotemark[3] & NA
      & 3.63\footnotemark[3] & 0.00 & 3.68\footnotemark[3] & NA \\
    \ch{FeO}\footnotemark[2] & 3.48 & 3.51\footnotemark[3] & -0.01 & 3.60
      & -0.02 & 3.65 & 0.00 & 3.69 & -0.01 \\
    \ch{CoO} & 2.45 & 2.51 & 0.01 & 2.61 & 0.01 & 2.64 & 0.01 & 2.65 & -0.02 \\
    \ch{NiO} & 1.37 & 1.46 & 0.00 & 1.61 & 0.01 & 1.56 & 0.00 & 1.60 & 0.00 \\
    \hline \hline
  \end{tabular*}
\footnotetext[1]{Solution with a band gap.}
\footnotetext[2]{Solution with no band gap.}
\footnotetext[3]{Predicted ground state.}
\end{table*}

Besides the electronic band structure, magnetism is another property
that can also be used to check the correctness of the implementation
of the MGGA potential. In a previous work \cite{TranPRB20}, the spin
magnetic moment $\mu_{S}$ of ferromagnetic (FM) and antiferromagnetic
(AFM) solids was calculated with numerous MGGA functionals. However,
due to the unavailability of the self-consistent MGGA implementation,
$\mu_{S}$ was calculated using the fixed spin-moment (FSM)
\cite{SchwarzJPF84} and $C$-shift methods for FM and AFM systems,
respectively. In both methods the total energy $E_{\text{tot}}$
[Eq.~(\ref{eq:Etot})] is minimized with respect to the magnitude of
the moment $\mu_{S}$ (which is varied within a certain range), and the
value of $\mu_{S}$ at the minimum of $E_{\text{tot}}$ is the value
that should in principle correspond to the value calculated
self-consistently. In Ref.~\cite{TranPRB20}, the orbitals were
calculated using the optimal GGA potentials (the same as those used in
the present work) and it was shown that in some cases it is of
importance to use the optimal potential (instead of the standard PBE)
in order to mitigate the error due to the non-self-consistent
procedure.

Here, we compare the magnetic moments from Ref.~\cite{TranPRB20} with
those obtained self-consistently. Such a comparison was already done
in Ref.~\cite{TranPRB20} for TPSS and SCAN in the case of MnO, FeO,
CoO, and NiO, however the self-consistent results were obtained with
other codes (VASP \cite{KressePRB96} and GPAW
\cite{EnkovaaraJPCM10,FerrighiJCP11}). Table~\ref{tab:moment} shows
the WIEN2k self-consistent values of $\mu_{S}$ for FM metals (Fe, Co,
and Ni) and AFM insulators (MnO, FeO, CoO, and NiO) systems, and the
value in parenthesis indicates the difference with respect to the
results obtained with the FSM/$C$-shift methods
~\cite{TranPRB20}. Note that the atomic magnetic moments on the
transition-metal atoms in the AFM solids are defined according to the
Bader volume from the quantum theory of atoms in molecules, as
implemented in the Critic2 code
\cite{OterodelaRozaCPC09,OterodelaRozaCPC14}. The agreement between
the two ways of calculating $\mu_{S}$ is excellent, which again should
demonstrate that the MGGA potential is implemented correctly into the
WIEN2k code. The largest difference, obtained for Fe with SCAN, is
only -0.03~$\mu_{\text{B}}$ (the negative sign indicates that the FSM
value from Ref.~\cite{TranPRB20} is larger). Compared to the $C$-shift
method the current self-consistent implementation has the large advantage
of not needing to rely on the additional variational procedure to find
the magnetic moments $\mu_s$ for each atomic sphere, a process that
would be especially complicated and expensive for a supercell
calculation (e.g. surface or a system with a defect).

In the particular case of FeO, an important point should be
noted. Depending on the functional, two different solutions can be
stabilized. They differ in the occupation of the Fe-$3d$ orbitals, one
corresponding to a metallic character and the other to a state with a
band gap. With LDA and most common GGAs only the metallic state can be
obtained. Both solutions can be obtained with GGAs having a large
enhancement factor like AK13 \cite{ArmientoPRL13} and (probably most)
MGGAs. Table~\ref{tab:moment} shows the FeO results for both
solutions. It is possible to obtain the metallic and non-metallic
states with the MGGAs in the case of self-consistent calculations. To
obtain the non-metallic state for FeO, we had to `manually' populate
the correct spin-down $a_{1g}$ instead of the $e_g'$ orbital (for
instance with a properly occupied density matrix in a GGA+U
calculation) to open the gap~\cite{mazinInsulatingGapFeO1997}, after
which the MGGAs would converge and correctly predict this state as the
ground state. When resuming a self-consistent cycle from a PBE
calculation or from a superposition of atomic densities, we obtain a
gapless state with all MGGAs considered here. This is not the case for
CoO, where the MGGAs would converge to the correct gapped state
independent of the starting point of the self-consistency
cycle. Therefore, only one solution can be found for AFM CoO.

Despite several attempts only the metallic state can be
obtained with the $C$-shift method for TPSS, SCAN, and TASK. This
example of FeO shows the limitations of using the FSM/$C$-shift method
instead of doing the self-consistent calculation. Finally, it is
interesting to note that the ground state is the non-metallic state
with SCAN, HLE17, and TASK, but the metallic one with TPSS. In the
case of CoO, all four MGGAs lead to a non-metallic ground state, while
PBE predicts a metallic ground state. HLE17 yields the largest gap of
$1.55$~eV, followed by SCAN ($1.16$~eV), TASK ($0.51$~eV), and TPSS
($0.42$~eV). All of these values are still smaller than the
experimental value of $2.5 \pm 0.3$~eV
\cite{vanelpElectronicStructureCoO1991}.

Finally, a few words should be said about the comparison with
experiment. As shown in Ref.~\cite{TranPRB20}, as well as in other
works
\cite{IsaacsPRM18,JanaJCP18a,EkholmPRB18,FuPRL18,FuPRB19,MejiaRodriguezPRB19},
SCAN and TASK lead to magnetic moments that are clearly larger than
experiment for FM metals, while TPSS is rather similar to PBE (i.e.,
very slight overestimation) and HLE17 is quite irregular. For the AFM
insulators, SCAN and TASK improve with respect to PBE, which strongly
underestimates the atomic magnetic moment. Again, TPSS is similar to
PBE, while HLE17 is irregular.

\section{Lattice constant: Effect of self-consistency}
\label{sec:latticeconstant}

\begin{table*}
  \caption{\label{tab:lattice}Equilibrium lattice constant (in \AA) of
    solids calculated self-consistently (SC) or non-self-consistently
    (NSC) with MGGA functionals. Indicated in parenthesis is the GGA
    potential that is used for calculating the core electrons and
    radial functions $f_{\mu\ell m}$ (for SC calculations) or for
    calculating the orbitals plugged into the MGGA functional (for NSC
    calculations). For comparison, the experimental values are 4.205
    (Na), 6.043 (Cs), 5.412 (Si), 4.905 (Pb), 3.599 (Cu), 5.569
    (NaCl), 5.640 (GaAs), 4.334 (FeO) , and 2.659 ($a$) and 4.863 ($c$)
    (Zn)
    \cite{belskyNewDevelopmentsInorganic2002,bergerhoffInorganicCrystalStructure1983,nussStructuralAnomalyZinc2010}.}
  \begin{tabular*}{\linewidth}{@{\extracolsep{\fill}}lSSSSSSSS}
    \hline \hline
      \multicolumn{1}{l}{} &
\multicolumn{4}{c}{{TPSS}} & \multicolumn{4}{c}{{SCAN}} \\
\cline{2-5}\cline{6-9} \multicolumn{1}{l}{} &
\multicolumn{1}{c}{{SC(RPBE)}} & \multicolumn{1}{c}{{SC(PBE)}} &
\multicolumn{1}{c}{{NSC(RPBE)}} & \multicolumn{1}{c}{{NSC(PBE)}} &
\multicolumn{1}{c}{{SC(RPBE)}} & \multicolumn{1}{c}{{SC(PBE)}} &
\multicolumn{1}{c}{{NSC(RPBE)}} & \multicolumn{1}{c}{{NSC(PBE)}} \\ \hline
  Na & 4.241 & 4.242 & 4.237 & 4.237 & 4.213 & 4.211 & 4.205 & 4.207 \\
  Cs & 6.280 & 6.279 & 6.275 & 6.264 & 6.238 & 6.235 & 6.230 & 6.226 \\
  Si & 5.458 & 5.458 & 5.457 & 5.457 & 5.435 & 5.436 & 5.433 & 5.435 \\
  Pb\footnotemark[1] & 4.983 & 4.983 & 4.984 & 4.984 & 4.972 & 4.972 & 4.978 & 4.974 \\
  Pb\footnotemark[2] & 4.983 & 4.983 & 4.983 & 4.983 & 4.976 & 4.976 & 4.979 & 4.975 \\
  Cu & 3.576 & 3.576 & 3.577 & 3.576 & 3.558 & 3.558 & 3.556 & 3.557 \\
  NaCl & 5.702 & 5.703 & 5.704 & 5.700 & 5.583 & 5.583 & 5.580 & 5.583 \\
  GaAs & 5.711 & 5.711 & 5.711 & 5.711 & 5.658 & 5.658 & 5.653 & 5.657\\
  FeO\footnotemark[3] & 4.281 & 4.281 & NA & NA & 4.274 & 4.273 & NA & NA \\
  FeO\footnotemark[4] & 4.274 & 4.274 & 4.274 & 4.274 & 4.259 & 4.259 & 4.268 & 4.269 \\
  Zn ($a$) & 2.639 & 2.639 & 2.640 & 2.638 & 2.574 & 2.574 & 2.576 & 2.574\\
  Zn ($c$) & 4.727 & 4.727 & 4.725 & 4.733 & 4.988 & 4.989 & 4.971 & 4.975\\
  \hline
  \multicolumn{1}{l}{} & \multicolumn{4}{c}{HLE17} &
\multicolumn{4}{c}{TASK} \\ \cline{2-5}\cline{6-9}
\multicolumn{1}{l}{} & \multicolumn{1}{c}{SC(mRPBE)} &
\multicolumn{1}{c}{SC(PBE)} & \multicolumn{1}{c}{NSC(mRPBE)} &
\multicolumn{1}{c}{NSC(PBE)} & \multicolumn{1}{c}{SC(HCTH/407)} &
\multicolumn{1}{c}{SC(PBE)} & \multicolumn{1}{c}{NSC(HCTH/407)} &
\multicolumn{1}{c}{NSC(PBE)} \\
  \hline
  Na & 4.049 & 4.050 & 4.044 & 4.048 & 4.740 & 4.741 & 4.617 & 4.624 \\
  Cs & 6.087 & 6.093 & 6.079 & 6.090 & 7.547 & 7.549 & 7.094 & 7.180 \\
  Si & 5.257 & 5.256 & 5.255 & 5.249 & 5.531 & 5.534 & 5.519 & 5.523 \\
  Pb\footnotemark[1] & 4.870 & 4.869 & 4.871 & 4.863 & 5.143 & 5.146 & 5.147 & 5.140 \\
  Pb\footnotemark[2] & 4.870 & 4.870 & 4.870 & 4.860 & 5.143 & 5.147 & 5.144 & 5.138 \\
  Cu & 3.510 & 3.510 & 3.510 & 3.485 & 3.601 & 3.597 & 3.600 & 3.591 \\
  NaCl & 5.575 & 5.570 & 5.574 & 5.540 & 6.273 & 6.272 & 6.188 & 6.267 \\
  GaAs & 5.553 & 5.552 & 5.552 & 5.540 & 5.755 & 5.760 & 5.750 & 5.747 \\
  FeO\footnotemark[3] & 4.169 & 4.166 & NA & NA & 4.383 & 4.393 & NA & NA \\
  FeO\footnotemark[4] & 4.151 & 4.148 & 4.149 & 4.150 & 4.365 & 4.374 & 4.362 & 4.360 \\
  Zn ($a$) & 2.584 & 2.588 & 2.584 & 2.578 & 2.617 & 2.615 & 2.607 & 2.605 \\
  Zn ($c$) & 4.655 & 4.687 & 4.655 & 4.627 & 5.131 & 5.131 & 5.131 & 5.125 \\
  \hline \hline
\end{tabular*}
\footnotetext[1]{With the $4f$ and $5s$ subshells treated
  in the core.}
\footnotetext[2]{With the $4f$ and $5s$ subshells treated
in the valence.}
\footnotetext[3]{Solution with a band gap.}
\footnotetext[4]{Solution with no band gap.}
\end{table*}

This section presents the results for the equilibrium lattice constant
of selected solids: Na, Cs, Si, Pb, Cu, NaCl, GaAs, FeO (both states
with and without band gap), ZnO and Zn. As in
Sec.~\ref{sec:bandgap} for the band gap, the purpose is to
illustrate the influence of the GGA potential for calculating the core
orbitals and radial functions $f_{\mu\ell m}$. The results, see columns `SC'
(self-consistent) in Table~\ref{tab:lattice}, show that using either
PBE or the corresponding optimal potential has very little
influence. The largest difference is found in the $c$ lattice constant
of Zn with HLE17, where the difference between SC(mRPBE) and SC(PBE)
reaches 0.03~\AA. This is an outlier, because the second largest
differences reach only around 0.005~\AA, and are obtained for Cs (with
HLE17), NaCl (with HLE17), GaAs (with TASK), and FeO (with TASK).
Thus, as in the case of the band gap, the choice of the GGA potential
for the core electrons and radial functions (either the standard PBE
or the variationally optimal one) is unimportant in the vast majority
of cases. Note that two sets of calculations were done for Pb: with
the $4f$ and $5s$ subshells treated either in the core (thus, with a
GGA potential) or in the valence (thus, with the MGGA potential). The
results are basically the same. This confirms again that the treatment
of the core electrons with a suitable GGA potential is a good
approximation.

Also shown in Table~\ref{tab:lattice}, are the equilibrium lattice
constants obtained non-self-consistently (`NSC' columns) using the
orbitals generated either from the PBE potential or the optimal GGA
one. For the insulating state of FeO, such a calculation is not
possible (see the discussion in Sec.~\ref{sec:magnetism}), and this is
indicated with 'NA' (not available) in the
Table~\ref{tab:lattice}. Such a NSC procedure for calculating the
lattice constant, bulk modulus, or cohesive energy with MGGA
functionals has been used in previous works (see, e.g.,
Refs.~\cite{TranJCP16,ZhangNJP18}). The GGA potential to generate the
orbitals has a larger influence in some cases. This is expected since
the GGA potential is applied to all electrons, and not only to the
core electrons as for SC calculations. The influence of the GGA
potential on the lattice constant reaches 0.08~\AA for Cs and NaCl,
both with TASK. With TPSS and SCAN the GGA potential has as a small
influence, while non-negligible differences are seen with HLE17.

To obtain these results for Cu with HLE17 (PBE), Zn with TASK (NSCF on
PBE and HCTH/407), FeO with HLE17 (both GGAs), it was necessary to
add HDLOs to the basis set, in order to
reduce the linearization error~\cite{KarsaiCPC17}. Without these
additional basis functions the WIEN2k code generated the same
warnings as described in Sec.~\ref{sec:bandgap}; the variational
energy was higher and the result was much further away from the proper
one obtained with HLE17 (mRPBE). This shows again that in some
cases, the addition of HDLOs can mitigate
the use of a less-than-optimal GGA potential for the construction of
the radial basis functions.

Comparing now the SC and NSC results, very large differences are
obtained in the case of the TASK functional. For Na, Cs, and NaCl, the
differences are about 0.1, 0.5, and 0.09~\AA, respectively. With TPSS
and SCAN, the differences between the NSC and SC results are very
small. In the vast majority of cases, using the orbitals generated
from the optimal GGA potential (instead of the PBE one) leads to the
best agreement with the SC results. Clear exceptions are Na, Cs, and NaCl
with TASK.

As a side comment, we mention that SCAN is one of the most accurate
functionals for the lattice constant of solids, while TPSS is slightly
less accurate and shows a (moderate) tendency to overestimate the
values \cite{TranJCP16,ZhangNJP18}. The two other functionals
  provide inconsistent results. While it is quite accurate in some
  cases (Zn and GaAs), HLE17 leads to some large underestimations (Na,
  Zn (c)) \cite{VermaJPCC17}, while TASK leads to huge
  overestimations (Na, Cs) as shown in Table~\ref{tab:lattice}.

\begin{table}[ht]
  \centering
  \caption{Comparison of various functionals to experiment for the
    $c/a$ ratio (in \AA) and the unit cell volume (in \AA$^{3}$) of
    hcp Zn. For the MGGA functionals the values of SCF calculations
    using the optimal GGA for core and radial functions are used.}
  \label{tab:zinc}
  \begin{tabular*}{\linewidth}{@{\extracolsep{\fill}}lccccc}
    \hline \hline
      & LDA & PBE & TPSS & SCAN & Expt.\footnotemark[1] \\
      \hline
    $c/a$ & 1.844 & 1.902 & 1.791 & 1.937 & 1.829 \\
    vol. & 27.12 & 30.49 & 28.51 & 28.62 & 29.77 \\
    \hline \hline
    \end{tabular*}
     \footnotetext[1]{Ref.~\cite{nussStructuralAnomalyZinc2010}}
\end{table}

For the particular case of Zn, it is interesting to compare the
results of the in general better-performing MGGAs (HLE17 and
TASK severely over- and underbind respectively) with the LDA and PBE
functionals in Table~\ref{tab:zinc}. Compared to LDA, all three
functionals improve upon the underestimation of the volume. However, both
PBE and SCAN sacrifice the accurately predicted $c/a$ ratio of
LDA. TPSS does yield more accurate values for both the $c/a$ ratio and the
unit cell volume.

We finish by comparing in Table~\ref{tab:lattice2} our SCAN lattice
constants with the results from Refs.~\cite{ZhangNJP18} and
\cite{MejiaRodriguezPRB18} that were obtained with the FHI-aims
\cite{BlumCPC09} (non-self-consistently using PBE orbitals) and VASP
(self-consistently) codes, respectively. The agreement with the
FHI-aims code is excellent, while sizable differences with VASP, in
particular for NaCl occur.

\begin{table}[ht]
\caption{\label{tab:lattice2}Comparison of the SCAN equilibrium
lattice constants (in \AA) calculated with different codes.}
\begin{tabular*}{\linewidth}{@{\extracolsep{\fill}}lSSS}
  \hline \hline
    {Solid} & {WIEN2k-SC(RPBE)} & {FHI-aims\footnotemark[1]} & {VASP\footnotemark[2]} \\
    \hline
    Na & 4.211 & 4.207 & 4.193 \\
    Cs & 6.235 & {NA} & 6.227 \\
    Si & 5.436 & 5.433 & 5.429 \\
    Cu & 3.558 & 3.558 & 3.566 \\
    NaCl & 5.583 & 5.585 & 5.563 \\
  GaAs & 5.658 & 5.656 & 5.659 \\
  \hline \hline
  \end{tabular*}
  \footnotetext[1]{Ref.~\cite{ZhangNJP18}.}
\footnotetext[2]{Ref.~\cite{MejiaRodriguezPRB18}.}
\end{table}

\section{Application: Electric field gradient}
\label{sec:EFG}
\begin{table*}
  \caption{\label{tab:EFG}EFG (in $10^{21}$ V/m$^{2}$) on a
    transition-metal atom in elemental metals and CuO, \ch{Cu2O}, and
    \ch{Cu2Mg} calculated with MGGA functionals at experimental
    lattice constants. For comparison, results from
    Ref.~\cite{TranPRM18} obtained with the PBE, GLLB-SC, and HSE06
    methods are also shown. The error bars of the experimental values
    are calculated from the uncertainty in the quadrupole moment and
    quadrupole coupling constants when available. The values which
    show large errors are underlined.}
  \sisetup{table-format = 2.2(2), table-text-alignment = center,
  table-number-alignment = center, table-column-width = 0pt}
 \begin{tabular*}{\linewidth}{@{\extracolsep{\fill}}l@{}SSSSSSSSS@{}}
  \hline \hline
   {Method} & \ch{Ti} & \ch{Zn} & \ch{Zr} & \ch{Tc} & \ch{Ru} & \ch{Cd}
   & \ch{CuO} & \ch{Cu2O} & \ch{Cu2Mg} \\
    \hline
   PBE & 1.73 & 3.49 & 4.19 & -1.61
   & \multicolumn{1}{l}{\underline{\tablenum[table-format=2.2]{ -1.46}}} & 7.54
   & \multicolumn{1}{l}{\underline{\tablenum[table-format=2.2]{ -2.83}}}
   & \multicolumn{1}{l}{\underline{\tablenum[table-format=2.2]{ -5.54}}} & -5.70 \\
   TPSS & 1.69 & 3.33 & 4.20 & -1.60
   & \multicolumn{1}{l}{\underline{\tablenum[table-format=2.2]{ -1.38}}} & 7.36
   & \multicolumn{1}{l}{\underline{\tablenum[table-format=2.2]{ -3.74}}}
   & \multicolumn{1}{l}{\underline{\tablenum[table-format=2.2]{ -5.60}}} & -5.69 \\
   SCAN & 1.75
   & \multicolumn{1}{l}{\underline{\tablenum[table-format=2.2]{4.38}}} & 4.37
   & -2.03 & \multicolumn{1}{l}{\underline{\tablenum[table-format=2.2]{-1.66}}}
   & \multicolumn{1}{l}{\underline{\tablenum[table-format=2.2]{9.47}}} & -7.15
   & \multicolumn{1}{l}{\underline{\tablenum[table-format=2.2]{ -6.41}}} & -5.70 \\
   HLE17 & 1.70 & 3.50
   & \multicolumn{1}{l}{\underline{\tablenum[table-format=2.2]{3.83}}}
   & \multicolumn{1}{l}{\underline{\tablenum[table-format=2.2]{ -0.93}}}
   & -0.73 & 7.44 & -6.33 & -8.59
   & \multicolumn{1}{l}{\underline{\tablenum[table-format=2.2]{ -4.76}}} \\
   TASK & 1.76
   & \multicolumn{1}{l}{\underline{\tablenum[table-format=2.2]{4.77}}} & 4.72
   & -1.66 & \multicolumn{1}{l}{\underline{\tablenum[table-format=2.2]{ -1.42}}}
   & \multicolumn{1}{l}{\underline{\tablenum[table-format=2.2]{10.33}}}
   & \multicolumn{1}{l}{\underline{\tablenum[table-format=2.2]{ -3.03}}} & -9.56 & -5.78 \\
   GLLB-SC & 1.62 & 3.72 & 4.42 & -1.66 & -1.26 & 8.05
   & \multicolumn{1}{l}{\underline{\tablenum[table-format=2.2]{-4.65}}}
   & -9.99 & -5.58 \\
   HSE06 & 1.5
   & \multicolumn{1}{l}{\underline{\tablenum[table-format=2.2]{4.4}}} & 4.5
   & -2.0 & \multicolumn{1}{l}{\underline{\tablenum[table-format=2.2]{-1.3}}}
   & \multicolumn{1}{l}{\underline{\tablenum[table-format=2.2]{9.4}}} & -8.9
   & -8.3 & \multicolumn{1}{l}{\underline{\tablenum[table-format=2.2]{-6.3}}} \\
   Expt.\footnotemark[1] & 1.57(12) & 3.40(35) & 4.39(15) & 1.83(9) & 0.97(11)
   & 7.60(75) & 7.55(52) & 10.08(69) & 5.76(39) \\
    \hline \hline
  \end{tabular*}
  \footnotetext[1]{Ref.~\cite{TranPRM18}.}
\end{table*}

As a final application of the MGGA functionals we discuss the
calculation of the EFG~\cite{BlahaPRB88}. We consider here the
transition-metal atom in the elemental metals and Cu-compounds listed
in Table~\ref{tab:EFG}. These are the same systems that we considered
in a previous work \cite{TranPRM18}, where we showed that among a
plethora of methods the GLLB-SC potential
\cite{GritsenkoPRA95,KuismaPRB10} leads overall to the best agreement
with experiment. The screened hybrid functional HSE06
\cite{HeydJCP03,KrukauJCP06} was shown to be also rather good overall
in comparison to the other methods. From the results shown in
Table~\ref{tab:EFG}, it is clear that GLLB-SC is still the most
accurate method for EFG calculations. It is close to the experimental
value in all but one case, which is CuO.

For all other methods (which includes both the GGAs and the MGGAs) the
results are mixed. We find for each of them three or four cases which
have a large disagreement with experiment. The errors are not very
systematic. For example, SCAN overestimates the EFG for Cd, but
underestimates the one of \ch{Cu2O}. One trend that can be observed,
is that SCAN always leads to similar or larger values for the EFG
compared to its `predecessors' PBE and TPSS. This suggests that the
ground-state density with SCAN has an increased asphericity.

CuO and \ch{Cu2O} are systems for which the standard LDA and PBE
functionals lead to qualitatively wrong results, with values that are
between two and four times smaller than experiment. It was shown
\cite{TranPRM18,TranPRB11} that HSE06 (for CuO and \ch{Cu2O}) and
GLLB-SC (for \ch{Cu2O}) substantially improve the results. A few other
semilocal methods also improve for \ch{Cu2O} \cite{TranPRM18}. The
results of the present work show that SCAN and TASK are accurate
only for one or the other (CuO for SCAN and \ch{Cu2O} for TASK), while
HLE17 leads to reasonable EFG values for both compounds.

\section{Summary}
\label{sec:summary}

In summary, the self-consistent implementation of MGGA functionals
into the WIEN2k code, which is based on an all-electron APW based method,
has been presented. The formalism has been discussed in detail, and a
comparison with results from the literature for the band gap of 30
solids shows very good agreement between the implementations. Magnetism has
also been considered, and the magnetic moments obtained
self-consistently are basically the same as those obtained
non-self-consistently with the FSM/$C$-shift method, thus again
showing that the new self-consistent implementation is reliable.

Then, the effect due to self-consistency on the lattice constant
revealed to be rather minor for the TPSS, SCAN, and HLE17 functionals,
but very large in some cases like Cs or NaCl with the TASK
functional. Finally, the EFG has been considered as an application. It
has been shown that some of the MGGAs are quite accurate for CuO and
\ch{Cu2O}, which are very difficult cases for standard GGAs.

A technical but rather important point concerned the GGA potential
that is used to calculate the core states and radial components of the
basis functions. A choice for the GGA potential has to be made and it
is recommended to use the one that is, for a given MGGA, the
variationally optimal one. However, we have shown that using the
standard PBE GGA potential leads to the same results in the vast
majority of cases.

Thus, this new implementation of MGGA functionals is accurate (since
implemented in an all-electron code) and leads to reliable results. It
uses the gKS scheme and is computationally barely more expensive
as other common semilocal methods like LDA or GGAs. Finally, we also
mention that the implementation has recently been used for the
calculation of the band gap of 2D materials in
Refs.~\cite{PatraJPCC21,TranJCP21}.

\begin{acknowledgments}

J.D. and P.B. acknowledge support from the Austrian Science Fund (FWF)
for Project W1243 (Solids4Fun). We are grateful to Miguel
A. L. Marques for useful discussions regarding Libxc and the VASP
calculations.

\end{acknowledgments}

\appendix

\section{Multiplicative part of the MGGA potential}
\label{sec:mult-part-mgga}

The developed expression of
$\nabla\cdot\left(\partial\epsilon_{xc}/\partial\nabla\rho_{\sigma}\right)$
in Eq.~(\ref{eq:vxc}) can be obtained by applying the chain
rule. Where $\xi=(\sigma,\sigma')$ is a shorthand index, it is given
by
\begin{align} \nabla\cdot\pdv{\epsilon_{xc}}{\nabla{\rho_{\sigma}}} &
= \sum\limits_{\sigma' \sigma''}^{\uparrow \downarrow} \left( 1 +
\delta_{\sigma \sigma'} \right) \pdv{\epsilon_{xc}}{\gamma_{\sigma
\sigma'}}{\rho_{\sigma''}}\gamma_{\sigma'\sigma''} \nonumber\\ &
+\sum\limits_{\sigma'}^{\uparrow \downarrow}
\sum\limits_{\xi}^{\uparrow\uparrow, \uparrow\downarrow, \downarrow
\downarrow} \left( 1 + \delta_{\sigma \sigma'}
\right)\pdv{\epsilon_{xc}}{\gamma_{\sigma \sigma'}}{\gamma_{\xi}}
\nabla{\rho_{\sigma'}} \cdot\nabla{\gamma_{\xi}} \nonumber\\ &
+\sum\limits_{\sigma'}^{\uparrow \downarrow} \left( 1 + \delta_{\sigma
\sigma'} \right)
\pdv{\epsilon_{xc}}{\gamma_{\sigma\sigma'}}\nabla^2{\rho_{\sigma'}}\nonumber\\
& +\sum\limits_{\sigma' \sigma''}^{\uparrow \downarrow} \left( 1 +
\delta_{\sigma \sigma'} \right)
\pdv{\epsilon_{xc}}{\gamma_{\sigma\sigma'}}{\tau_{\sigma''}}\nabla{\rho_{\sigma'}}\cdot\nabla{\tau_{\sigma''}}
\label{eq:gradexc}
\end{align}
where $\delta_{\sigma\sigma'}$ is the Kronecker delta.
$\gamma_{\sigma\sigma'} = \nabla \rho_{\sigma} \cdot \nabla
\rho_{\sigma'}$ is the contracted density gradient that needs to be
provided to Libxc \cite{MarquesCPC12,LehtolaSX18}, along with the
electron density $\rho_{\sigma}$ and the KED $\tau_{\sigma}$. The
output provided by Libxc is the exchange-correlation energy density
$\epsilon_{xc}$ and its (partial) derivatives. Note that due to
the $\tau_{\sigma}$-dependency of $\epsilon_{xc}$, there is an
additional term (the last one in Eq.~(\ref{eq:gradexc})) compared to
the GGA case.

\section{Derivation of the matrix element
$\mel**{\phi_{\mu}}{\hat{v}_{\tau}}{\phi_{\nu}}$}
\label{sec:derivation}

The KED-derived matrix element is evaluated using integration by
parts. When $\phi_{\nu}$ is an APW+lo basis function a surface term
must be included due to the discontinuity of the gradient $\nabla
\phi_{\nu}$ across the sphere boundary \cite{SjostedtSSC00}.
\begin{equation}
  \begin{split}
\label{eq:mel-partial-integrated}
\mel**{\phi_{\mu}}{\hat{v}_{\tau}}{\phi_{\nu}} = & \frac{1}{2} \left[
\sum\limits_{\alpha} \int\limits_{S_{\alpha}} +
\int\limits_{\mathrm{I}} \right] v_{\eta}\nabla{\phi}_{\mu}^{*} \cdot
\nabla\phi_{\nu}\dd^{3}\mathbf{r} {} \\ & \quad -\frac{1}{2}
\sum\limits_{\alpha} \oint\limits_{\partial S_{\alpha}} v_{\eta}
\phi_{\mu}^{*} \left( \nabla \phi_{\nu} \cdot \vu{r} \right) \dd \Omega
\end{split}
\end{equation} We consider the spherical term (volume integral over
the sphere $S_{\alpha}$), the surface term (integral over the sphere
boundary $\partial S_{\alpha}$) and the interstitial term separately.

\subsection{Spherical term}
\label{sec:volume-term}

In the spheres all quantities like the basis functions and the
potentials are expanded in spherical harmonics, allowing for a
separation of variables. The gradient of the basis functions
$\phi_{\mu} = \sum_{\ell{}m} f_{\mu \ell m} {Y}_{\ell m}$ is most
conveniently expressed using vector spherical harmonics
\cite{Varshalovich88,Arfken13}:
\begin{equation}
  \label{eq:grad_phi}
  \begin{split} \grad \phi_{\mu} = \sum\limits_{\ell{}m}
\sqrt{\frac{\ell}{2\ell + 1}} \left\{ \pdv{r} + \frac{\ell+1}{r}
\right\} f_{\mu \ell m} \mathbf{Y}_{\ell m}^{\ell - 1} {} \\
-\sqrt{\frac{\ell + 1}{2\ell + 1}} \left\{ \pdv{r} - \frac{\ell}{r}
\right\} f_{\mu \ell m} \mathbf{Y}_{\ell m }^{\ell +1}.
\end{split}
\end{equation}

The vector spherical harmonics can be defined in different bases; for
our purpose the spherical basis
\begin{align} \vu{e}_{+1} & = \frac{-1}{\sqrt{2}} \left( \vu{e}_x + i
\vu{e}_y \right)\\ \vu{e}_0 & = \vu{e}_z\\ \vu{e}_{-1} & =
\frac{1}{\sqrt{2}} \left( \vu{e}_x - i \vu{e}_y \right)
\end{align} is convenient. The vector spherical harmonics are then
\begin{equation}
\label{eq:VSH} \mathbf{Y}_{J M}^L = \sum\limits_{\gamma}^{-1, 0,
1}C_{L\: M-\gamma\: 1\: \gamma}^{J\: M} {Y}_{L M-\gamma}
\vu{e}_{\gamma},
\end{equation}
where $C_{\alpha \: \beta \: \gamma \: \delta}^{\epsilon \:
  \zeta}$ are Clebsch-Gordan coefficients. The dot product of two vector spherical harmonics with
differing quantum numbers is given by
\begin{equation}
  \label{eq:dot-prod-vsh}
  \begin{split} \mathbf{Y}_{J_1 M_1}^{*L_1} \cdot \mathbf{Y}_{J_2
M_2}^{L_2} = \sum\limits_{\mu} & C_{L_1\: M_1-\mu\: 1\: {\mu}}^{J_1\:
M_1} C_{L_2\: M_2-\mu\: 1\: {\mu}}^{J_2\: M_2} {} \\ & \times\quad
{Y}_{L_1 M_1-\mu}^{*} {Y}_{L_2 M_2-\mu}.
\end{split}
\end{equation}

Then, the dot product of the gradients of two
basis functions of Eq.~(\ref{eq:grad_phi}) can be partitioned in four
terms each involving a dot product of vector spherical harmonics,
coefficients involving $\ell$ and $m$, and radial integrals with four
contributions. The angular integrals of a product of three spherical
harmonics can be performed analytically and are given by the Gaunt
coefficients $G^{L\:M}_{\ell_1\: m_1\: \ell_2\: m_2}$:
\begin{equation} G^{L\:M}_{\ell_1\: m_1\: \ell_2\: m_2} = \int {Y}_{L
M}^{*}{Y}_{\ell_1 m_1} {Y}_{\ell_2 m_2} \dd\Omega.
\end{equation}

The four associated radial parts have each four terms. These are
performed numerically, and are given by
\begin{widetext}
\begin{align}
\label{eq:r--} R_{--} & = I_{L M} \big[ f_1^{*\prime} f'_2 r^2\big] +
\left( \ell_1 + 1 \right) I_{L M} \big[ f^{*}_1 f'_2 r\big] + \left(
\ell_2 + 1 \right) I_{L M} \big[ f_1^{*\prime} f_2 r \big] + \left(
\ell_1 + 1 \right) \left( \ell_2 +1 \right) I_{L M} \big[ f^{*}_1f_
2\big]\\
\label{eq:r-+} R_{-+} & = I_{L M} \big[ f_1^{*\prime} f'_2 r^2\big] +
\left( \ell_1 + 1 \right) I_{L M} \big[ f^{*}_1 f'_2 r\big] - \ell_2
I_{L M} \big[ f_1^{*\prime} f_2 r \big] - \ell_2 \left( \ell_1 + 1
\right) I_{L M} \big[ f^{*}_1f_2 \big] \\
  \label{eq:r+-} R_{+-} & = I_{L M} \big[ f_1^{*\prime} f'_2 r^2\big]
- \ell_1 I_{L M} \big[ f^{*}_1 f'_2 r\big] + \left( \ell_2 + 1 \right)
I_{L M} \big[ f_1^{*\prime} f_2 r \big] - \ell_1 \left( \ell_2 + 1
\right) I_{L M} \big[ f^{*}_1f_2 \big] \\
  \label{eq:r++} R_{++} & = I_{L M} \big[ f_1^{*\prime} f'_2 r^2\big]
- \ell_1 I_{L M} \big[ f^{*}_1 f'_2 r \big] - \ell_2 I_{L M} \big[
f_1^{*\prime} f_2 r \big] + \ell_1 \ell_2 I_{L M} \big[ f^{*}_1f_2
\big],
\end{align}
\end{widetext}
where $f_1 = f_{\mu \ell_1 m_1}$, $f_2 = f_{\nu \ell_2
m_2}$, and $I_{L M}$ is the integrated product with the angular
component of the KED-derived part of the potential.
\begin{equation}
\label{eq:1} I_{L M} \left[ g \right] =
\int\limits_{0}^{R_{\text{at}}}\left(v_{\eta}\right)_{LM}\left( r
\right) g \left( r \right) \dd r.
\end{equation}
The parenthesis $\left( \ldots \right)_{LM}$ indicate the $L M$
component of the spherical harmonics expansion, i.e.
$v_{\eta} = \sum\limits_{LM} \left( v_{\eta} \right)_{LM} Y_{LM}$

Putting together the four radial and angular parts with their
coefficients results in the following expression, for a single sphere:
\begin{widetext}
\begin{align}
\begin{split}
  \label{eq:mel-volume} \int\limits_S v_{\eta}\nabla{\phi}_{\mu}^{*}
\cdot \nabla\phi_{\nu}\dd^{3}\mathbf{r} = \sum\limits_{\left\{ \xi
\right\}} & \frac{1}{\sqrt{\left( 2\ell_1 +1 \right) \left( 2\ell_2 +1
\right)}} \\ & \times\left( \sqrt{\ell_1\ell_2} R_{--} C_{\ell_1-1\:
m_1-\gamma\: 1\: { \gamma}}^{\ell_1\: m_1} C_{\ell_2-1\: m_2-\gamma\:
1\: { \gamma}}^{\ell_2\: m_2} G_{L\: M\: \ell_2-1\:
m_2-\gamma}^{\ell_1-1\: m_1-\gamma} \right. \\ & - \sqrt{\ell_1 \left(
\ell_2 + 1 \right)} R_{-+} C_{\ell_1-1\: m_1-\gamma\: 1\: {
\gamma}}^{\ell_1\: m_1} C_{\ell_2+1\: m_2-\gamma\: 1\: {
\gamma}}^{\ell_2\: m_2} G_{L\: M\: \ell_2+1\: m_2-\gamma}^{\ell_1-1\:
m_1-\gamma}\\ & - \sqrt{\ell_2 \left( \ell_1 + 1 \right)}R_{+-}
C_{\ell_1+1\: m_1-\gamma\: 1\: { \gamma}}^{\ell_1\: m_1} C_{\ell_2-1\:
m_2-\gamma\: 1\: { \gamma}}^{\ell_2\: m_2} G_{L\: M\: \ell_2-1\:
m_2-\gamma}^{\ell_1+1\: m_1-\gamma} \\ & \left. + \sqrt{\left( \ell_1
+ 1 \right)\left( \ell_2 + 1 \right)} R_{++} C_{\ell_1+1\:
m_1-\gamma\: 1\: { \gamma}}^{\ell_1\: m_1} C_{\ell_2+1\: m_2-\gamma\:
1\: { \gamma}}^{\ell_2\: m_2}G_{L\: M\: \ell_2+1\:
m_2-\gamma}^{\ell_1+1\: m_1-\gamma} \right)
\end{split}
\end{align}
\end{widetext}
To obtain the result for all atoms in the unit cell, this expression
is multiplied by a phase factor
$e^{i \left( \mathbf{K'} - \mathbf{K}\right) \cdot
  \mathbf{R}_{\alpha}}$ and summed over the atomic indices $\alpha$.

The spherical terms are implemented similarly to the non-spherical
corrections in APW based methods. The main points are these. A list of
quantum numbers obeying the Gaunt selection rules should be
constructed, otherwise the loop over six $\ell, m$ combinations
becomes too expensive. For the potential lattice spherical harmonics
are used to exploit the point symmetry of the atomic sites. Secondly,
one should note that the radial integrals $R_{\pm \pm}$ are only
dependent on the azimuthal numbers of the basis functions $\ell_1$ and
$\ell_2$; the $m$-dependent factor enters only through the matching
coefficients $A^{\mathbf{K}}_{\ell m}$, $B^{\mathbf{K}}_{\ell m}$
which do not depend on the radial coordinate.

\subsection{Interstitial term}
\label{sec:interstitial-term}

The interstitial term  is much simpler to evaluate
since the gradient of the basis function is simply given by $\nabla
\phi_{\mathbf{K}} = i \mathbf{K} \phi_{\mathbf{K}}$. The result is
then
\begin{align}
  \label{eq:mel-interstitial}
  \begin{split} & \int\limits_{\mathrm{I}}
v_{\eta}\nabla{\phi}_{\mu}^{*} \cdot \nabla\phi_{\nu}\dd^{3}\mathbf{r}
\\ & \quad = \frac{1}{\Omega} \int\limits_{\text{cell}} \Theta \left(
\mathbf{r} \right) v_{\eta} \left( \mathbf{r} \right) \mathbf{K} \cdot
\mathbf{K'} e^{i \left( \mathbf{K'} - \mathbf{K} \right)\cdot
\mathbf{r}} \dd^3 \mathbf{r}
\end{split} \nonumber\\ & \quad = \mathbf{K} \cdot \mathbf{K'} \left(
\Theta v_{\eta} \right)_{\mathbf{G'} - \mathbf{G}},
\end{align}
the parenthesis $\left( \ldots \right)_{\mathbf{G}}$
indicate the fourier expansion coefficients, and
\begin{align} \Theta \left( \mathbf{r} \right) & =
\begin{cases} 0, &\mathbf{r} \in \text{S}_{\alpha} \\ 1, &\mathbf{r}
\in \text{I},
\end{cases}
\end{align}
is the step function. The notation
$\left( \theta v_{\eta} \right)_{\mathbf{G'} - \mathbf{G}}$ indicates
that the step function is multiplied in direct space, avoiding an
expensive convolution sum in reciprocal space \cite{Singh}.

\subsection{Surface term}
\label{sec:surface-term} The surface term is nonzero for APW(+lo)
basis functions with a discontinuous gradient across the sphere
boundary. Inside the sphere one has
\begin{align}
  \lim\limits_{r\rightarrow R_{\mathrm{at}}^{-}} \nabla \phi_{\mu} \left( \mathbf{r} \right) \cdot \vu{r} & =
  \sum\limits_{\ell m} f'_{\mu \ell m} Y_{\ell m} \bigg\rvert_{r = R_{\mathrm{at}}},
\end{align}
whereas in the interstitial, by using
the Rayleigh expansion,
\begin{align}
\lim\limits_{r\rightarrow R_{\mathrm{at}}^{+}} & \nabla \phi_{\mu} \left( \mathbf{r} \right) \cdot \vu{r} = \nonumber \\
  & \sum\limits_{\ell m} 4\pi i^l j'_l \left( K r \right)
Y^{*}_{\ell m} \left( \vu{k} \right) Y_{\ell m}\left( \vu{r} \right)
\bigg\rvert_{r = R_{\mathrm{at}}},
\end{align}
where the prime indicates the radial derivative. We can then define
the radial part of the expansion as

\begin{equation}
g_{\nu \ell m} \left( r \right) = 4\pi i^l j'_l \left( K_{\nu} r
\right) Y^{*}_{\ell m} \left( \vu{k_{\nu}} \right).\label{eq:2}
\end{equation}
The resulting
surface integral can then be performed analytically:
\begin{widetext}
\begin{align}
\label{eq:mel-surface}
  \begin{split} \oint\limits_{\partial S_{\alpha}} & v_{\eta}
\phi^{*}_{\mu} \left( \nabla \phi_{\nu} \cdot \vu{r} \right) \dd
\Omega = R^{2}_{\mathrm{at}} \sum\limits_{ \left\{ \xi \right\}}
\left( v_{\eta} \right)_{L M} G^{\ell_{1}\:m_1}_{L\: M\: \ell_2\: m_2}
\bigg( f^{*}_{\mu \ell_1 m_1} f'_{\nu \ell_2 m_2} - g^{*}_{\mu \ell_1
m_1} g'_{\nu \ell_2 m_2} \bigg) \Bigg\rvert_{r = R_{\mathrm{at}} }
   \end{split}
\end{align}
\end{widetext} where we used the spherical harmonics addition theorem,
$\sum\limits_{\left\{ \xi \right\}}$ signifies the sum over all
angular numbers. Note that the surface term is not Hermitian. In the
implementation it is explicitly made Hermitian.

The surface term is generally very small, and has a negligible effect
on all the calculations considered in this work. We verified this by
comparing well-converged LAPW and APW+lo calculations, and noting that
the difference in all cases was negligible. For the vast majority of
cases, the same is true when comparing APW+lo calculations including
or excluding the surface term.

We have noticed that in a few cases numerical problems arise with our
implementation of the surface term. We could not point out a single
cause for these problems, but they occur for the functionals with a
larger MGGA enhancement factor (like HLE17 or TASK) when using a large
plane-wave cutoff. Cases with a larger asphericity in the density and
potential (like Si) are also more strongly affected. Additionally, due
to the nature of the APW basis functions, the potential itself becomes
discontinuous. This occurs in both the multiplicative and the
non-multiplicative parts and is a consequence of the functions only
being matched in value and not in slope at the sphere boundary.
Choosing smaller sphere sizes also helped mitigate these numerical
issues.

Because calculations excluding the surface term are numerically more
reliable, we chose not to include this surface term in the APW+lo
calculations for the present work.

\bibliography{collection}

\end{document}